\documentclass[superscriptaddress,aps,pre,onecolumn, 12pt]{revtex4-2}

\usepackage[utf8]{inputenc}
\usepackage{amsmath}
\usepackage{amsfonts}
\usepackage{amssymb}
\usepackage{graphicx}
\usepackage[toc,page]{appendix}
\usepackage[colorlinks]{hyperref}

\hypersetup{colorlinks=true,
	linkcolor=blue,
	anchorcolor=black,
	citecolor=red,
	urlcolor=black
}

\usepackage[capitalise]{cleveref}
\creflabelformat{equation}{#2{\upshape(#1)}#3}

\providecommand{\nn}{\nonumber}
\providecommand{\bw}{\begin{widetext}}
\providecommand{\ew}{\end{widetext}}
\providecommand{\be}{\begin{equation}}
\providecommand{\ee}{\end{equation}}
\providecommand{\bal}{\begin{ailgn}}
\providecommand{\eal}{\end{align}}
\providecommand{\bea}{\begin{eqnarray}}
\providecommand{\eea}{\end{eqnarray}}
\providecommand{\beas}{\begin{eqnarray*}}
\providecommand{\eeas}{\end{eqnarray*}}

\begin{document}
 
\title{On non-ideal chemical-reaction networks and phase separation}

\author{Ander Movilla Miangolarra}
\altaffiliation{To whom correspondence should be addressed: ander.movilla-miangolarra@jic.ac.uk}
\affiliation{Laboratoire Physico-Chimie Curie, Institut Curie, PSL Research University, CNRS UMR 168, Paris, France}
\affiliation{Sorbonne Universit\'es, UPMC Univ. Paris 06, Paris, France} 
\affiliation{Present address: Computational and Systems Biology Department, John Innes Centre, Norwich, United Kingdom}
\author{Michele Castellana}
\affiliation{Laboratoire Physico-Chimie Curie, Institut Curie, PSL Research University, CNRS UMR 168, Paris, France}
\affiliation{Sorbonne Universit\'es, UPMC Univ. Paris 06, Paris, France}


\begin{abstract}
Much of  the theory on chemical-reaction networks (CRNs) has been developed in the ideal-solution limit, where interactions between the solutes are negligible. However, there is a large variety  of phenomena in biological cells and soft-matter physics which appear to deviate from the ideal-solution behaviour.
Particularly striking is the case of liquid-liquid phase separation, which is typically caused by inter-particle interactions. Here, we revisit a number of  known results in the domain of ideal CRNs, and  we generalise and adapt them to arbitrary interactions between the solutes which stem from a given free energy. 
Among these is the form of the steady-state probability distribution and Lyapunov functions for complex-balanced networks, where the creation and  annihilation rates are equal for all chemical complexes which appear as reactants or products in the CRN. 
Finally, we draw a phase diagram for complex-balanced reaction-diffusion solutions based on the minimisation of such Lyapunov function with a rationale similar to that of equilibrium thermodynamics, but for systems that may sustain non-equilibrium chemical currents at steady state.
Nevertheless, we find that complex-balanced networks are not sufficient to create diffusion currents at steady state.
\end{abstract}

\keywords{Statistical physics, nonequilibrium thermodynamics, stochastic systems}
\maketitle

\section{Introduction}

The cytoplasm of a cell does not behave like an ideal solution \cite{Zielinski2017Nonideal}, since, in many cases, interactions among the solutes cannot be neglected. Indeed, in the cytoplasm there is a plethora of interactions among  proteins, other macromolecules, and ions. 
Some of the most common interactions that are relevant in the cellular cytoplasm are steric and crowding effects \cite{Mitchison2019ColloidCrowding,Minton2008Crowding}, as well as electrostatic interactions \cite{Fall2001MitochondrialCa2+,Hyman2018Grammar}. Arguably, the most striking phenomenon caused by these interactions is the emergence of phase-separated condensates, also known as membraneless organelles in the cell-biology literature, which are now widely studied \cite{brangwynne2009germline,Rosen2012PhaseTrans,Vale2016PhaseSep}. The composition of these membraneless organelles is different from the one of the cytoplasm, because they are typically enriched in a specific type of molecules while they exclude others  \cite{Rosen2018Whosin}. Moreover, it has been hypothesised that such organelles spatially control biochemical reactions, by modulating their rates and specificity within the condensate \cite{Banani2017Review,castellana2014enzyme,vagne2015sensing,buchner2013optimization}.

Given their  important role in the internal spatial organisation of
cells, the regulation of phase-separation phenomena is crucial for
many cellular functions. One of the ways in which cells can
dynamically control the onset, composition and function of
membraneless organelles is through chemical reactions, notably
post-translational modifications  like phosphorylation
\cite{FormanKay2019Phosphoregulated,FormanKay2019Phosphodependent} or
methylation  \cite{FormanKay2016Nuage}. However, phase separation is
also triggered by changes in the environment
\cite{Hyman2018PhYeast,Rosen2021Functions}, thus establishing
biological condensates as potential switch-like sensing and regulatory mechanisms.

While most of the insights outlined above are the result of extensive experimental efforts,  the interplay between interactions within the solution and non-equilibrium chemical reactions has also been widely studied from the theoretical standpoint. 
Most of these efforts \cite{Huberman1976striations,Glotzer1995reaction,Li2020noneq,FanLee2018CRCPhase} have been based on effective reaction-diffusion models that can describe patterning and non-equilibrium phenomena in a simple way, but lack thermodynamic consistency. More precisely, in these approaches the reaction dynamics is modelled with mass-action kinetics (MAK), which implicitly assumes that the solution is ideal (while the interaction-influenced diffusion that drives phase-separation is not), which leads to the aforementioned lack of consistency.
Early progress in reconciling the spatial patterns predicted by these models with a thermodynamically consistent description  was limited to a linear-stability analysis of binary systems \cite{LefeverCarati1997}. More recently, some works aimed at establishing a deterministic theory for non-ideal chemical-reaction networks (CRNs)  \cite{Esposito2021NonidealCRN}, the relation between phase coexistence and chemical kinetics \cite{bauermann2021chemical}, and exploring minimal examples for pattern formation with non-ideal CRNs \cite{Bazant2013,Zwicker2021controlling}.  Nevertheless, the link between non-equilibrium CRNs and phase separation has not yet been  elucidated in full generality.
 
Here, we aim at building a thermodynamically consistent framework for interacting reaction-diffusion systems which may exhibit phase-separation at steady state. 
Therefore, in this framework, in the same way diffusion is governed by a free energy (that takes into account the interactions), the dynamics of the chemical reactions must also reflect this free-energetic dependency. 
Here, previous efforts are complemented by analysing the behaviour of non-ideal CRNs in the stochastic limit, in an effort to build a complete theory. 
This explicit description of non-ideality in the CRN allows us to naturally adapt and generalise the results from the well-established theory of ideal CRNs. We do so by first constructing a framework such that, in the absence of explicit non-equilibrium driving, the system relaxes to thermodynamical equilibrium. Then, we focus on complex-balanced networks, for which the steady-state creation and annihilation rate of each chemical complex are equal. For these type of CRNs, we derive the steady-state probability distributions and Lyapunov functionals, which allows us to obtain the steady-state concentration profiles. 

The paper is organised as follows: In Section \ref{sec_crn} we describe the dynamics of spatially homogeneous CRNs in the stochastic  and deterministic limit, introduce the concept of complex balance, and recall the main features of MAK. In Section \ref{stoch_dynamics}, we impose  thermodynamical constraints on the reaction rates for CRNs at equilibrium, by consistently relating these rates to the free energy, and discuss how they can be modified in non-equilibrium settings.
In Section \ref{sec_claim} we generalise to non-ideal CRNs the known result for the steady-state distribution of complex-balanced networks. Building on this result, in Section \ref{sec_lyap}  we propose a candidate Lyapunov function of complex-balanced systems.
In the same Section, we generalise the previous Lyapunov function to systems with spatial inhomogeneities, and derive the resulting phase diagram for a non-equilibrium, complex-balanced, chemically reactive mixture. Finally, in Section \ref{disc} we discuss the interpretations and implications of our results. 

\section{Chemical-Reaction Networks}\label{sec_crn}

A chemical-reaction network (CRN) is composed of $N$ chemical species and $M$ reaction pathways; which we assume reversible, for a better alignment with thermodynamic principles. A reaction within the CRN, denoted by the label $\rho$, is specified as follows:
\begin{equation}
\sum_a r^{\rho}_a X_a \leftrightharpoons \sum_a s^{\rho}_a X_a,
\end{equation}
where $X_a$, $a=1, \cdots, N$ is one the $N$ species in the network. In the rest of this paper, the indexes $a$ and $b$ will be used for chemical species only. 

The matrices $r^{\rho}_a$ and $s^{\rho}_a$  denote the 
number of particles of each species participating in the forward and backward reaction, respectively, i.e., $r^{\rho}_a$ specifies the number of reactants of type $a$ in the forward reaction $\rho$, and $s^{\rho}_a$, that of the products of type $a$ in the backward reaction. 
Note that, given that the reactions are taken to be reversible, the distinction between reactants and products is arbitrary. 

The amount of particles of species $a$ created along the forward reaction is denoted by
 \begin{equation}\label{eq_v}
v^\rho_a=s^{\rho}_a-r^{\rho}_a.
\end{equation}
We also introduce the vectors $\bm{v}^\rho= (v^\rho_1, \cdots, v^\rho_N) $, $\bm{r}^\rho=( r^\rho_1 \cdots r^\rho_N)$ and $\bm{s}^\rho=( s^\rho_1, \cdots, s^\rho_N)$.
, and the matrices 
\be\label{eq_V}
\bm{V} =\,  
\left(\begin{array}{c}
\vdots\\
\bm{v}^\rho\\
\vdots
\end{array}\right), \,
 \\
\bm{R} = \left(\begin{array}{c}
\vdots\\
\bm{r}^\rho\\
\vdots
\end{array}\right)
,\,
\bm{S}= \, \left(\begin{array}{c}
\vdots\\
\bm{s}^\rho\\
\vdots
\end{array}\right)
.
\ee
Finally, we define a complex $\bm{z}$ as the number and type of particles that participate in a chemical reaction as either reactants ($\bm{z}=\bm{r}^\rho$) or products ($\bm{z}=\bm{s}^\rho$). A single complex $\bm{z}$ may appear in more than one reaction within the network. In order to clarify the definitions above, we illustrate them for the following network. 

\subsection{Example}\label{ex1}
 For the CRN
\begin{align}
\label{CRN1}
\rm{A}+\rm{B} &\leftrightharpoons \rm{C},\\\label{CRN2}
 \rm{B} &\leftrightharpoons \rm{D},
\end{align}
the vectors of reactants are 
\begin{equation}
\bm{r}^1=
\begin{pmatrix}
1, & 1, & 0, & 0  
\end{pmatrix},\,
\bm{r}^2=
\begin{pmatrix}
0, & 1, & 0, & 0
\end{pmatrix},\,
\end{equation}
and the product vectors are
\begin{equation}
\bm{s}^1=
\begin{pmatrix}
0, & 0, & 1, &  0
\end{pmatrix},\,
\bm{s}^2=
\begin{pmatrix}
0, &  0, & 0, & 1
\end{pmatrix} ,
\end{equation}
where each of the entries in the vector  correspond to different species and the reactions are labelled by the superindices. Finally, we have the vectors
\begin{equation}
\bm{v}^1=
\begin{pmatrix}
-1, & -1, & 1, &  0
\end{pmatrix},\,
\bm{v}^2=
\begin{pmatrix}
0, &  -1, & 0, & 1
\end{pmatrix} ,
\end{equation}
that specify the net amount of particles of each species created by each of the reactions occurring once in the forward direction.
\vspace{0.5cm}

Building on the definitions above, in what follows we introduce the stochastic and deterministic description of a CRN, the starting point of the rest of this work. 

\subsection{Stochastic description}

If the chemical species in the solution diffuse fast (with respect to the typical timescale of chemical reactions) and is stirred regularly, the system may be considered to be well mixed and it can be described in terms of a single homogeneous concentration of each of the species across space. 
Then, a state of the system---the number of particles of each type---is determined by the vector 
\begin{equation}
\bm{n}=( n_1, \cdots, n_N).
\end{equation}
Each state $\bm{n}$  of the system has a probability measure $P(\bm{n}, t)$ at any instant of time $t$. The dynamics for the probability of states of homogeneous CRNs is given by the Chemical Master Equation (CME), which reads \cite{Gillespie1992rigurous}:

\begin{equation}
\label{CME}
\frac{\partial P(\bm{n}, t)}{\partial t}= \sum_\rho f_{+\rho} (\bm{n}-\bm{v}^{\rho}) P(\bm{n}-\bm{v}^{\rho}) + \sum_\rho f_{-\rho} (\bm{n}+\bm{v}^{\rho}) P(\bm{n}+\bm{v}^{\rho})- \sum_\rho [f_{+\rho} (\bm{n})+f_{-\rho} (\bm{n}) ]P(\bm{n}) ,
\end{equation}
where the summations over $\rho$ run over all reactions in the CRN, $\bm{v}$ is given by \cref{eq_v}, and the rate of the transitions in the network is given by the \textit{propensity functions} $f_{\pm\rho}$, $+\rho$ corresponding to the forward direction of the reaction and $-\rho$ corresponding to the backward direction. \cref{CME} is different to other statements of the CME because we made explicit the fact that every reaction is reversible.

\subsection{Deterministic description}\label{det_desc}

For large particle numbers, by averaging both sides of \cref{CME} and assuming vanishing correlations according to the mean-field picture---assumptions that are supposed to be accurate in the large-particle number limit---one can work out a set of equations for the concentrations $c_a = n_a/V$ in the macroscopic limit, where  both $n_a$ and $V$ are large. 
In this limit, the state of the system is specified by the concentrations $c_1, \cdots, c_N$, and one obtains the following classical set of equations for the dynamics of  the concentrations in a CRN 
\cite{Grima2017Approximation}:
\begin{equation}
\label{deterministic_kinetics}
\frac{\partial c_a}{\partial t} = \sum_\rho v_a^\rho (J_{+\rho}-J_{-\rho}),
\end{equation}
where the currents $J$ still need to be determined. 
While both the deterministic and stochastic descriptions refer to the same system, the former one is only accurate for large particle numbers, also known as the thermodynamic limit, where fluctuations are negligible.

\subsection{Complex balance}

\begin{figure}
\centering
\includegraphics[width=0.4\textwidth]{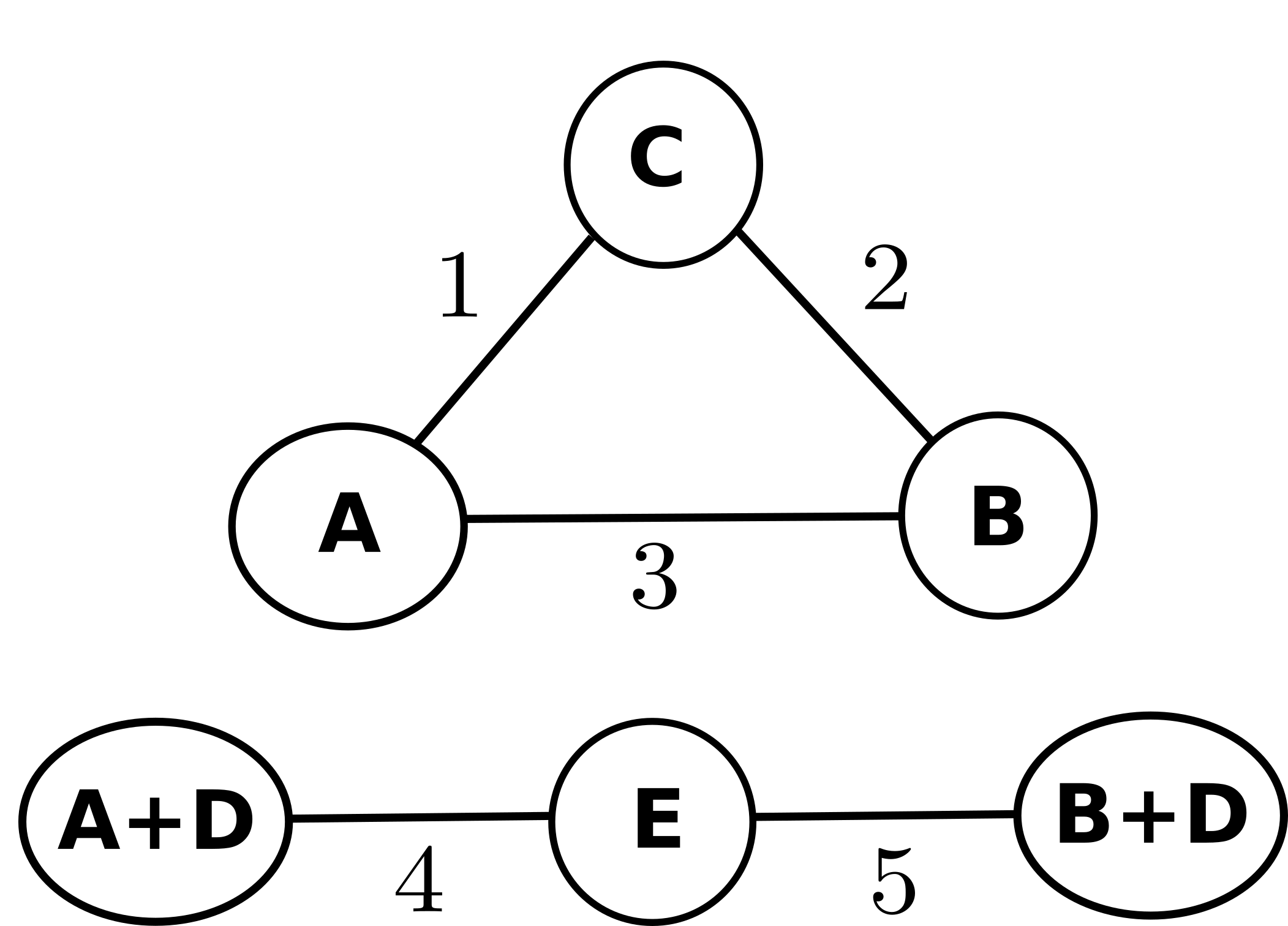}
\caption{\label{fig_ex_hierarchy} Graphical representation of a chemical-reaction network (CRN). The CRN has six complexes, where each is represented as a node in the graph: $\rm A$, $\rm B$, $\rm C$, $\rm{A}+\rm{D}$, $\rm E$ and $\rm{B}+\rm{D}$. The five reactions present in the CRN are numbered.}
\end{figure}

In a CRN, a \textit{complex} is a  set of chemical species and their respective particle numbers, which take part in a reaction, as either reactants or  products. Its most general expression is the vector
\begin{equation}
\bm{z}=(z_1, \cdots, z_a, \cdots ),
\end{equation}
where the index $a$ runs over all chemical species and the integer $z_a$ is the number of molecules of the species $X_a$ in the complex $\bm{z}$. 
Any CRN can be represented as a graph whose nodes denote the complexes that take part in the reactions, where there is an edge between two complexes if and only if there is a  reaction $\bm{z}^m \leftrightharpoons \bm{z}^n$ in the CRN, and $\bm{z}^m$ and $\bm{z}^n$  are two different complexes---see for example Fig \ref{fig_ex_hierarchy}.

In a deterministic CRN,  whose kinetics are given by \cref{deterministic_kinetics}, the \textit{creation  rate of the complex} $\bm{z}$, $J_{+\bm{z}}$,  is defined as
\begin{equation}\label{j_complex}
J_{+\bm{z}}=\sum_{\rho \vert \bm{s}^\rho=\bm{z}} J_{+\rho} +\sum_{\rho \vert \bm{r}^\rho=\bm{z}} J_{-\rho}
\end{equation}
where both addends in \cref{j_complex} are source terms for the complex $\bm{z}$:  The subscript $\rho \vert \bm{s}^\rho=\bm{z}$ indicates that the sum is taken over the reactions $\rho $ whose product complex equals the complex $\bm{z}$, and, similarly, $\rho \vert \bm{r}^\rho=\bm{z}$ indicates that that the sum is taken over those reactions whose reactant complex equals $\bm{z}$. Proceeding along the same lines, we define the \textit{rate of annihilation of the complex} $\bm{z}$ as
\begin{equation}
J_{-\bm{z}}=\sum_{\rho \vert \bm{s}^\rho=\bm{z}} J_{-\rho} +\sum_{\rho \vert \bm{r}^\rho=\bm{z}} J_{+\rho}.
\end{equation}

A deterministic network is said to have a \textit{complex-balanced steady state} if its steady state satisfies the condition that the creation rate and the annihilation rate of each complex  are equal \cite{Horn1972}:
\begin{equation}\label{complex_bal}
J_{+\bm{z}}=J_{-\bm{z}} \quad \;\;\; \forall \, \bm{z}.
\end{equation}
While there exist some topological conditions in the CRN which ensure that the steady state is complex balanced \cite{feinberg1972complex}, not every CRN possesses a complex-balanced non-equilibrium steady-state. We refer the interested reader to Refs. \cite{Feinberg1995,Anderson2010ProductFormCRN,Polettini2015Deficiency} for more detailed discussions on the topological constraints that determine complex-balancing and its consequences for MAK networks.   \\


In general, a complex-balanced steady state is one of many steady states which a CRN may have. We can order these types of steady states in terms of their generality as follows: 

\begin{figure}
\centering
\includegraphics[width=\textwidth]{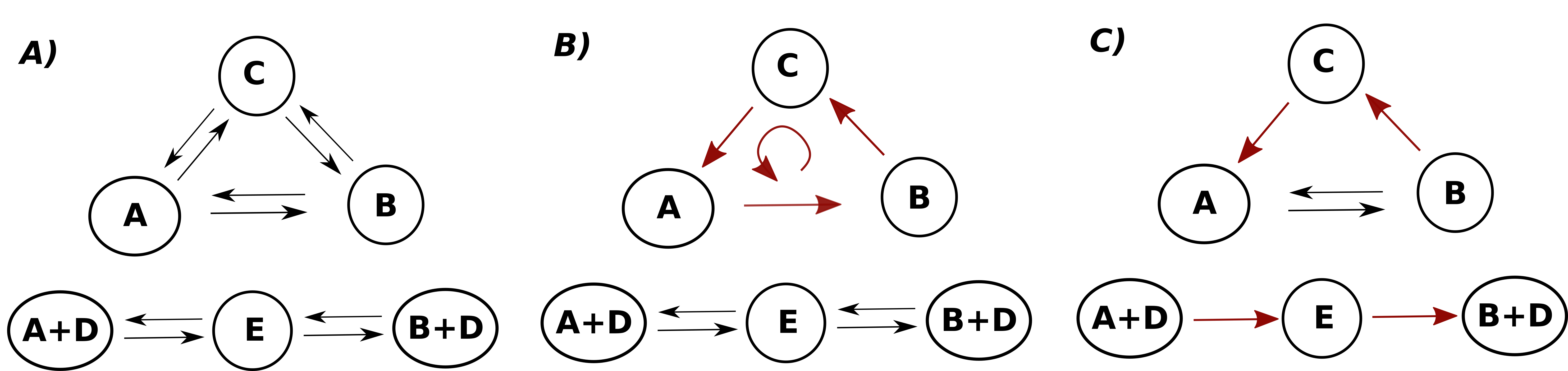}
\caption{Graphical depiction of the different types of steady states, taking the network of Fig. \ref{fig_ex_hierarchy} as example. A) Detailed-balanced steady state. Here, the backward and forward rates of each reaction are equal. B) Complex-balanced steady state. There exist non-vanishing net currents (red arrows) but the steady state still satisfies the complex-balanced requirement, \cref{complex_bal}. It only allows for cycles at steady state that can be visualised from the network in terms of complexes (red cycle). C) General steady state. There exist non-vanishing net currents (red arrows) and the complex-balanced requirement does not hold.}
\label{fig_ex_levels} 
\end{figure}

\begin{description}
\item[Equilibrium steady state] The most restrictive condition that we can impose to a steady state is detailed balance:
\begin{equation}
\label{H_DB}
J_{+\rho}=J_{-\rho} \quad \;\;\; \forall \, \rho.
\end{equation}
\Cref{H_DB} corresponds to a system at thermodynamic equilibrium, and implies that the rate of the forward reaction equals the rate of the backward reaction for every reaction $\rho$ in the CRN (example depicted in Fig. \ref{fig_ex_levels} A). \\

\item[Complex-balanced steady state] More general than detailed-balanced steady states are complex-balanced steady states, which satisfy 
\begin{equation}
\label{H_CB}
\sum_{\rho \vert \bm{s}^\rho=\bm{z}} J_{+\rho} +\sum_{\rho \vert \bm{r}^\rho=\bm{z}} J_{-\rho}=\sum_{\rho \vert \bm{s}^\rho=\bm{z}} J_{-\rho} +\sum_{\rho \vert \bm{r}^\rho=\bm{z}} J_{+\rho} \quad \;\;\; \forall \, \bm{z}.
\end{equation}
\Cref{H_CB} implies that the creation and the annihilation rate of each complex  $\bm{z}$ are equal. In the example of Fig. \ref{fig_ex_hierarchy}, the only way to have a complex-balanced steady state which is not at equilibrium (i.e. is not entirely detailed-balanced) is by taking the net rate $J_{+\rho}-J_{-\rho}$ in reaction 1 equal to  the net rate of those of reactions 2 and 3, and thus reactions 4 and 5 must be detailed balanced (since the system has to be at steady state). This steady state is depicted in Fig. \ref{fig_ex_levels} B. \\

\item[General steady state] The most general class of steady states is defined by the vanishing time derivatives of the dynamical equation (\ref{deterministic_kinetics}). By splitting the contributions of each complex $\bm{z}$, this condition can be rewritten as
\begin{equation}
\label{H_GSS}
\sum_m z_a^m \Bigg(\sum_{\rho \vert \bm{s}^\rho=\bm{z}_m} J_{+\rho} +\sum_{\rho \vert \bm{r}^\rho=\bm{z}_m} J_{-\rho}\Bigg)  = 
\sum_m z_a^m \Bigg(\sum_{\rho \vert \bm{r}^\rho=\bm{z}_m} J_{+\rho} +\sum_{\rho \vert \bm{s}^\rho=\bm{z}_m} J_{-\rho}\Bigg)  \;\;\;\forall \, a ,
\end{equation}
where $m$ is an index that labels each of the complexes in the network and the integer $z_a^m$ represents its components. 
As a result, there are no constraints between the net rates of each reaction other than those imposed by the stationarity condition of \cref{deterministic_kinetics}. 
In the example of Fig. \ref{fig_ex_hierarchy}, this implies that, at steady state, there can be current cycles where, for example, species $\rm{A}$ is created by reaction 1, but annihilated by reaction 4 through the complex $\rm{A+D}$, thus breaking complex balance (see Fig. \ref{fig_ex_levels} C). 
\end{description}

From this hierarchical classification, it can be clearly seen that detailed balance, \cref{H_DB}, implies complex balance, \cref{H_CB}, which, in turn, implies the steady-state condition, \cref{H_GSS}. However, the converse is not true: A general steady state is not necessarily  complex balanced, and a complex-balanced steady state  is not necessarily detail balanced. Therefore, complex balance is less restrictive of a constraint than detailed balance, but it is more restrictive than a generic steady state.

\subsection{Complex balance in networks with mass-action kinetics} \label{complex}

As a particular instance of special importance, in what follows we will discuss complex balance in ideal CRNs with 
MAK.

In short, MAK consists of the hypothesis that the rate of the chemical reaction is proportional to the product of the concentrations of the reactants: As a result, in the deterministic description,  the MAK expressions for the currents read
\begin{equation}
\label{MAK_det}
J_{+\rho}=k_{+\rho} \prod_a c_a^{r_a^\rho}, \quad J_{-\rho}=k_{-\rho} \prod_a c_a^{s_a^\rho},
\end{equation}
where  $k_{\pm \rho}$ are the rate constants. In what follows, we will denote by $c^\ast_a$ the steady-state concentration of species $a$ in the deterministic description.  Then, for a deterministic system with MAK, the complex-balance condition is given by 
\begin{equation}
\label{CB_MAK}
\sum_{\rho \vert \bm{s}^\rho=\bm{z}}  k_{+\rho} \prod_a (c_a^\ast)^{r_a^\rho} +\sum_{\rho \vert \bm{r}^\rho=\bm{z}} k_{-\rho} \prod_a (c_a^\ast)^{s_a^\rho}  =\sum_{\rho \vert \bm{s}^\rho=\bm{z}} k_{-\rho} \prod_a (c_a^\ast)^{s_a^\rho} +\sum_{\rho \vert \bm{r}^\rho=\bm{z}} k_{+\rho} \prod_a (c_a^\ast)^{r_a^\rho} \quad \; \forall \, \bm{z}.
\end{equation}




Conversely, in the stochastic description [with dynamics is given by \cref{CME}] 
the MAK expressions for $f_{\pm\rho}$  become 
\begin{align}
\label{stoc_mass_action1}
f_{+\rho} (\bm{n})= & k_{+\rho} \prod_a \frac{n_a!}{(n_a-r_a^{\rho})!},\\\label{stoc_mass_action2}
f_{-\rho} (\bm{n})= &  k_{-\rho} \prod_a \frac{n_a!}{(n_a-s_a^{\rho})!}.
\end{align}

Previous studies  \cite{Anderson2010ProductFormCRN} have shown that the steady state of complex-balanced CRNs with MAK is known to have a product-form expression in terms of independent Poisson distributions, and it reads
\begin{equation}
\label{ideal_cb_ness}
\pi (\bm{n})=\prod_{a=1}^M \frac{(c^{\ast}_a)^{n_a}}{n_a!}e^{-c^{\ast}_a},
\end{equation}
linking the deterministic steady state ($c_a^\ast$) to the stochastic steady state $\pi$.
Furthermore, in Ref. \cite{Anderson2010ProductFormCRN} it is shown that, if the propensity functions take the more general form
\begin{align}
\label{theta_rates}
f_{+\rho} (\bm{n})=& k_{+\rho} \frac{\theta (\bm{n}) }{\theta (\bm{n}-\bm{r}^\rho) },\\
f_{-\rho} (\bm{n})=& k_{-\rho} \frac{\theta (\bm{n}) }{\theta (\bm{n}-\bm{s}^\rho) },
\end{align}
then the steady-state distribution reads 
\begin{equation}
\label{Anderson_gen}
 \pi (\bm{n})= \frac{M}{\theta (\bm{n})}\prod_{a=1} (c^\ast_a)^{n_a}, 
\end{equation}
where $M$ is a normalisation constant and $\theta$ a function which maps the vector of integer numbers $\bm{n}$ into a real-valued positive number. 

In what follows, we will  demonstrate that the result \eqref{ideal_cb_ness} can be generalised to the non-ideal case, i.e., to a class of propensity functions $f$ which take into account the  physical interactions between molecules.

\section{Reaction rates for non-ideal chemical-reaction networks}\label{stoch_dynamics}

In the previous Section we introduced the general description of CRNs, both on a stochastic and deterministic level: in either cases, a choice for the propensity functions, or currents, must be made to set the network dynamics. For ideal solutions, the most common choice is MAK, as outlined above. However, in what follows we consider solutes which mutually interact and which are, therefore, not ideal, and specify the propensity functions.

\subsection{Equilibrium systems}\label{sec_eq}

Here, we consider CRNs at thermodynamic equilibrium, i.e., systems which are not subject to external, non-equilibrium driving. Given that the system is an equilibrium one, at steady state the principle of detailed balance must hold for every reaction $\rho$: The probability flux across a reaction $\rho$ in the forward direction must equal the one in the backward direction. In this Section, we will impose the detailed-balance condition on the propensity functions at thermal equilibrium and suggest a generalisation for systems out of equilibrium. 

\subsubsection{Stochastic description}\label{sec_stoc_eq}
In the stochastic description, the detailed-balance condition at steady state reads
\begin{equation}
\label{detailed_balance_peq}
P^{\textrm{eq}}(\bm{n})f_{+\rho}(\bm{n})=f_{-\rho} (\bm{n}+\bm{v}^\rho) P^{\textrm{eq}}(\bm{n}+\bm{v}^\rho),
\end{equation} where  the equilibrium probability distribution $P^{\textrm{eq}}(\bm{n})$ for  closed stochastic systems---total number of particles fixed---is given by the canonical Boltzmann distribution:
\begin{equation}
\label{canonical}
P^{\textrm{eq}}(\bm{n})=\frac{1}{Z}e^{-\beta F(\bm{n})},
\end{equation} 
with $\beta = 1/(k_{\rm B}T)$, $k_{\rm B}$ is the Boltzmann constant, $T$ the temperature, $ F(\bm{n})$ the Helmholtz free energy of the system in state $\bm{n}$, and $Z$ a normalisation factor---the partition function in statistical physics. For systems that exchange mass with a single particle reservoir, the equilibrium distribution (\ref{canonical}) is replaced by the distribution for the grand-canonical ensemble \cite{Rao2018Conservation}.

Combined with \cref{canonical}, the detailed-balance condition in \cref{detailed_balance_peq} yields the following constraint for the propensity functions:
\begin{align}
\label{rates_peq}
\frac{f_{+\rho}(\bm{n})}{f_{-\rho} (\bm{n}+\bm{v}^\rho)} =& \frac{P^{\textrm{eq}}(\bm{n}+\bm{v}^\rho)}{ P^{\textrm{eq}}(\bm{n})}\\ \nn
=&e^{-\beta[ F(\bm{n}+\bm{v}^\rho)- F(\bm{n})]}.
\end{align}
Then, we choose the following functional form for the propensity functions:
\begin{align}
\label{gen_rate+}
f_{+\rho}(\bm{n})=\, &k_\rho e^{\beta[ F(\bm{n})-F(\bm{n}-\bm{r}^{\rho})]}, \\ \label{gen_rate-}
f_{-\rho} (\bm{n})=\, &k_\rho e^{\beta [F(\bm{n})-F(\bm{n}-\bm{s}^{\rho})]},
\end{align}
where $k_\rho$ is the reaction constant, which needs to be equal in both the forward and the backward reaction for \cref{rates_peq} to be satisfied. 
Given that  the free energy $F$  may, in general, depend on the inter-particle interactions---such as steric, electrostatic, or other  interactions---\cref{rates_peq} implies that the chemical-reaction rates may depend on these inter-particle interactions.

The choices \eqref{gen_rate+} and \eqref{gen_rate-} for the propensity functions are not unique, but it is particularly appealing because it reduces to MAK for ideal systems. In fact, consider an ideal lattice-model solution with $\mathcal{N}$ particles including both solvent and solute---see Appendix \ref{virial_rates} for details. The free energy is
\begin{equation}
\label{Fid}
F_{\textrm{id}}=\sum_a n_a \mu_a^0 +\frac{1}{\beta}\Bigg[\sum_a \log (n_a !) - \log (\mathcal{N}!)\Bigg],
\end{equation}
where $\mathcal{N}=\sum_a n_a$ (including solvent particles in the sum) and $\mu_a^0$ is the 
standard-state chemical potential of species $a$ (taken with respect a given reference state noted as `0'), which may depend on parameters like temperature or nature of the solvent and the solute $a$. Then, the rates take the following form:
\begin{align}
\label{DF_mass_action}
f_{+\rho}(\bm{n})=&k_\rho e^{\beta[ F(\bm{n})-F(\bm{n}-\bm{r}^{\rho})]}\\ \nn
=&k_\rho e^{\beta \sum_a r^{\rho}_a \mu_i^0} \frac{(\mathcal{N}-\sum_a r_a^\rho)!}{\mathcal{N}!} \prod_a \frac{n_a!}{(n_a-r_a^{\rho})!},
\end{align}
where $(\mathcal{N}-\sum_a r_a^\rho)!/\mathcal{N}!$ can be approximated by $\mathcal{N}^{-\sum_a r_a^\rho}$. Setting
\begin{equation}
k_{+\rho}=k_\rho \frac{\exp(\beta \sum_a r^{\rho}_a \mu_a^0)}{\mathcal{N}^{\sum_a r_a^\rho}},
\end{equation} 
we obtain that $f_{+\rho}$ coincides with the MAK propensity function \eqref{stoc_mass_action1}, and similarly for $f_{-\rho}$ and \cref{stoc_mass_action2}. \vspace{0.75cm}\\

We conclude this Section with a remark on the reaction constant, $k_\rho$: In \cref{gen_rate+,gen_rate-} we have assumed that $k_\rho$ is a constant of the reaction, independent on the state $\bm n$ of the system. However, in general $k_\rho$ may depend on $\bm{n}$, because the system itself is part of the environment where the chemical reactions take place.  These effects can be disregarded for most cases in ideal solutions (since they are usually dilute), but they may not be negligible in non-ideal systems. For instance, in the case of phase separation, the multiple phases of the system may constitute very different environments for the chemical reactions, accelerating them or slowing them down.

Independently of whether $k_\rho$ in \cref{gen_rate+,gen_rate-} depends on the system state or not,   detailed balance, Eq. (\ref{rates_peq}), must still hold. This means that the forward reaction constant for a state $\bm{n}$ must be equal to the backward reaction constant for a state $\bm{n}+\bm{v}^\rho$. One way to ensure this equality while keeping the state-dependency of the reaction constants, is to make $k_\rho$ a function of the state $\bm n$ deprived of the reactant complex, i.e., $\bm{n}-\bm{r}^\rho$, for the forward case, and of $\bm{n}+\bm{v}^\rho-\bm{s}^\rho$ for the backward one:
\begin{align}
\label{gGMAK_neq+}
f_{+\rho}(\bm{n})=&k_\rho(\bm{n}-\bm{r}^\rho) e^{\beta[ F(\bm{n})-F(\bm{n}-\bm{r}^{\rho})]}, \\ \label{gGMAK_neq-}
 f_{-\rho} (\bm{n}+\bm{v}^\rho)=&k_\rho(\bm{n}+\bm{v}^\rho-\bm{s}^\rho) e^{\beta [F(\bm{n}+\bm{v}^\rho)-F(\bm{n}+\bm{v}^\rho -\bm{s}^{\rho})]},
\end{align}
 where \cref{gGMAK_neq+,gGMAK_neq-} satisfy \cref{rates_peq} because 
\begin{equation}
\bm{n}-\bm{r}^\rho=\bm{n}+\bm{v}^\rho-\bm{s}^\rho,\end{equation} 
see \cref{eq_v}.

\begin{figure}
\centering
\includegraphics[scale=0.2]{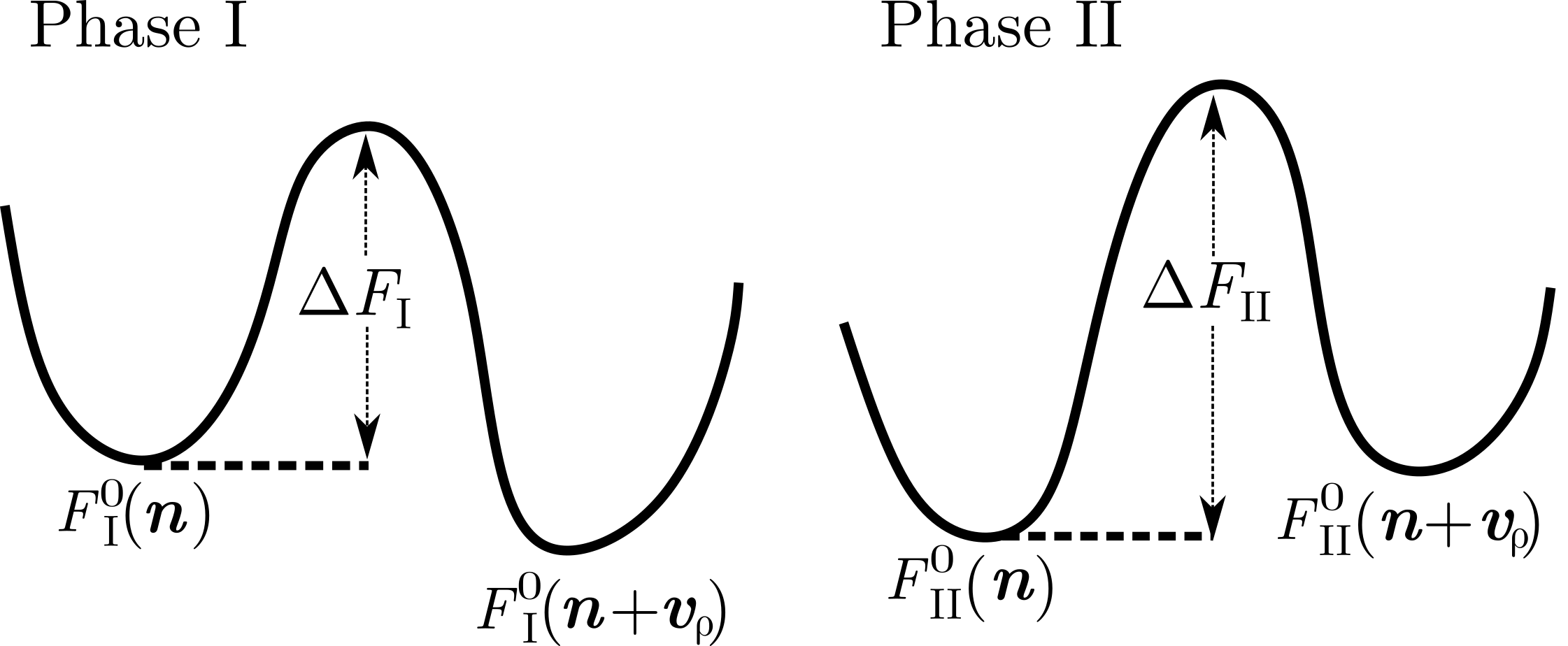}
\caption{Free-energy landscape for a chemical reaction. The horizontal dimension is the reaction coordinate and the vertical one specifies the height of the reaction free energy $F^0$---the free energy $F$ with the entropic term 
$\sum_a\log n_a!$ removed \cite{prigogineThermodynamics}.
The quantities $F^0(\bm{n})$ and $F^0(\bm{n}+\bm{v}_\rho)$ denote the free energy of the system before and after the reaction, 
subscripts $\rm I$ and $\rm II$ specify the system phase, and 
$\Delta F$ the height of the free-energy barrier. 
 For each phase, there are two minima in the free-energy landscape, corresponding to whether the reaction has occurred or not, see the left- and right-hand minimum, respectively.  In this example, $\Delta F$ depends on the phase the reaction takes place in, and, thus, the reaction constant $k_\rho$ would also depend on the environment in which the reaction occurs.
}
\label{F_landscape}
\end{figure}

The dependency above of $k_\rho$ on the system state can be pictured as follows. In analogy with the classical transition-state theory, we can think of the microscopic mechanism of a  reaction as a random walk in a free-energy landscape \cite{HanggiReactionRate, prigogineThermodynamics}, see Fig. \ref{F_landscape}. Then, the value of the rate constant $k_\rho$ depends on the height of the free-energy barrier $\Delta F$ of the reaction. While the free energies of reactants and products (the stable local minima in the reaction landscape) have free energies $F^0_{\rm I}, F^0_{\rm II}$ defined by $F$, this is not the case for the barrier height $\Delta F$. The dependency of the height of the barrier---and  thus of $k_\rho$---on the system state is precisely the one discussed in \cref{gGMAK_neq+,gGMAK_neq-}, and it may strongly affect the CRN dynamics.  In summary, we are connecting the chemical reaction rates to the free energy of the system $F$, but also to  $\Delta F$ which sets the value of the reaction constants $k_\rho$.

\subsubsection{Deterministic description}\label{det_dynamics}

When the particle numbers $\bm{n}$ are large compared to the number of reactants and products, $\bm{r}$ and $\bm{s}$, respectively, the free-energy differences  which appear in the rates \eqref{gen_rate+} and \eqref{gen_rate-} can be rewritten as
\begin{align}
\label{approx_det}
F(\bm{n})-F(\bm{n}-\bm{r}^\rho) \approx &F(\bm{n})-\Bigg[F(\bm{n}) - \sum_a \frac{\partial F}{\partial n_a} r^\rho_a\Bigg]\\ \nn
=&  \sum_a r_a^\rho \mu_a,
\end{align}
where in the first line we expanded $F$ to first order in $\bm{r}$, in the second line we used the definition of the chemical potential of species $a$:
\begin{align}\label{eq_mu}
\mu_a=&\frac{\partial F}{\partial n_a} \\ \nn
 =& \frac{\partial \mathfrak{f}}{\partial c_a}, 
\end{align} 
and $\mathfrak{f}(\bm{c})$ is the free energy per unit volume in the deterministic notation. 

Therefore, the currents in a deterministic, non-ideal CRN
can be written as
\begin{align}
\label{det_rates}
J_{+\rho}=&k_\rho e^{\beta \sum_a r_a^\rho \mu_a},\\\nn
 J_{-\rho} =&k_\rho e^{\beta \sum_a s_a^\rho \mu_a},
\end{align}
which is an expression conceptually similar to that given by other approaches to construct thermodynamically consistent dynamics for deterministic CRNs \cite{Esposito2021NonidealCRN,Bazant2013}. Once again, the currents (\ref{det_rates}) match their ideal MAK counterpart (\ref{MAK_det}) if the chemical potentials used in the rates are those of an ideal solution, i.e., $\mu_a=1/ \beta\log c_a+\mu_a^0$. Here and in the rest of the text, dimensional arguments of the logarithms remain due to the fact that we are absorbing the effect of the total concentration in $\mu^0$, i.e., the original chemical potential was $\mu_a=1/ \beta\log (c_a/c_{\textrm{tot}})+\mu_a^0$, where $c_{\textrm{tot}}=\sum_a c_a$ (the sum includes the solvent), but since variations in $c_{\textrm{tot}}$ can be neglected $-\log (c_{\textrm{tot}})$ is just a constant and is absorbed into $\mu^0_a$ (and, thus, into  $k_{\rho}$).

As in the previous section, if we assume the rate constant is state-dependent then the currents are given by 
\begin{align}
\label{gen_rates_det}
J_{+\rho}(\bm{c})&=\tilde{k}_\rho g_\rho(\bm{c}) e^{\beta \sum_a r^\rho_a \mu_a},  \\ \nn
 J_{-\rho}(\bm{c})&=\tilde{k}_\rho g_\rho(\bm{c}) e^{\beta \sum_a s^\rho_a \mu_a},
\end{align} 
where $\tilde{k}_\rho$ is still a constant and any dependency of the rate constant on the state is given by the function $g_\rho(\bm{c})$.

\subsection{Non-equilibrium systems}\label{sec_non_eq}

So far we considered the propensity functions of equilibrium CRNs. Given the large number of physically interesting systems which are out of equilibrium, such as living beings, in what follows we will generalise the analysis of Section \ref{sec_eq} to a specific type of non-equilibrium systems: those in which the work is done by the chemostats they are connected to. 

Let us assume that $N'$ out of the $N$ species in the system are connected to multiple particle reservoirs---chemostats: In the stochastic and deterministic description, each chemostat keeps constant the chemical potential of the species to which it is connected. Then, in general, the system will not relax to equilibrium, because of the work done on it by the chemostats. In the stochastic and deterministic description, the dimensions of the space of states or concentrations, respectively, is reduced to $N-N'\leq N$, since the chemostatted species are no longer dynamical variables.

\subsubsection{Stochastic description}

We assume that connecting the system to several chemostats does not alter any of the mechanisms of the chemical reactions, since it only tunes the concentration of the species to which they are connected, in order to match a given value of chemical potential. Then, reactions that involve both chemostatted and non-chemostatted species are driven in one direction by the work done by the chemostats inserting and removing particles from the system (in order to keep their chemical potentials constant). Given that the mechanism of reaction remains the same, in line with the previous section the rates of these driven chemical transitions are taken to be
\begin{align}
\label{gen_rate_neq+}
f_{+\rho}(\bm{n})=\, &k_\rho e^{\beta[ F(\bm{n})-F(\bm{n}-\bm{r}^{\rho})+\sum_b r^\rho_b \mu_b]}, \\ \label{gen_rate_neq-}
f_{-\rho} (\bm{n})=\, &k_\rho e^{\beta [F(\bm{n})-F(\bm{n}-\bm{s}^{\rho})+\sum_b s^\rho_b \mu_b]},
\end{align}
where now $F$ is the free energy of the $N-N'$  non-chemostatted species, $\bm{n}$ contains the particle numbers of the non-chemostatted species only, and the summation over $b$ runs over the chemostatted species. 
For the sake of clarity, in what follows we will reserve the index $b$ for the chemostatted species, and the index $a$ for the non-chemostatted ones.

The rationale behind these relations is that the chemical reaction is still driven by free energy differences except that now the the free energy differences due to the consumption of chemostatted species is just given by the chemical potential of the chemostats $\mu_b$. The  terms $\sum_b r^\rho_b \mu_b$ and $\sum_b s^\rho_b \mu_b$ in the exponential represents the chemical work done by the chemostats (with chemical potentials fixed at $  \mu_b  $) when a reaction $\rho$ occurs,  which pushes the system out of equilibrium. The effect of the non-chemostatted species is still given by the free energy differences $F(\bm{n})-F(\bm{n}-\bm{r}^{\rho})$ and $F(\bm{n})-F(\bm{n}-\bm{s}^{\rho})$.

This implicitly assumes that the chemostatted species are abundant (so that the chemical potential does not fluctuate) and that they are ideal (negligible interactions with the non-chemostatted species). If the chemostatted species were not ideal, then the concentration of species might dynamically vary to match the chemostatted chemical potential as the particle numbers in the system change.  Here, we only consider the simpler case of ideal chemostatted species and refer the interested reader to Ref. \cite{Esposito2021NonidealCRN}, where the case of non-ideal chemostatted species was analysed.

As in Section \ref{sec_eq} [see \cref{gGMAK_neq+,gGMAK_neq-}], the rate constants in \cref{gen_rate_neq+,gen_rate_neq-} may be generalised in such a way that $k_\rho$ depends  on the system state: 

\begin{align}
\label{gen_noneq_rates}
f_{+\rho}(\bm{n})&=\tilde{k}_\rho g_\rho(\bm{n}-\bm{r}^\rho) e^{\beta[ F(\bm{n})-F(\bm{n}-\bm{r}^{\rho})+\sum_b r^\rho_b \mu_b]},  \\ \nn 
f_{-\rho} (\bm{n}+\bm{v}^\rho)&=\tilde{k}_\rho g_\rho(\bm{n}+\bm{v}^\rho -\bm{s}^{\rho}) e^{\beta[F(\bm{n}+\bm{v}^\rho)-F(\bm{n}+\bm{v}^\rho -\bm{s}^{\rho})+\sum_b s^\rho_b \mu_b]},
\end{align} 
where $\tilde{k}_\rho$ is independent of $\bm n$. Propensity functions of this form have been suggested before in other contexts, such as in the modelling of molecular motors \cite{julicher1997motors}.

\subsubsection{Deterministic description}\label{det_dynamics_neq}

Proceeding along the lines of Section \ref{det_dynamics}, in  the deterministic limit the above propensity functions result in the currents
\begin{align}
\label{gen_noneq_rates_det}
J_{+\rho}(\bm{c})&=\tilde{k}_\rho g_\rho(\bm{c}) e^{\beta( \sum_a r^\rho_a \mu_a+\sum_b r^\rho_b \mu_b)},  \\ \nn
 J_{-\rho}(\bm{c})&=\tilde{k}_\rho g_\rho(\bm{c}) e^{\beta( \sum_a s^\rho_a \mu_a+\sum_b s^\rho_b \mu_b)}. 
\end{align}

\section{Steady-state distribution for complex-balanced, non-ideal CRNs}\label{sec_claim}

In what follows, we will prove one of the central results of this work, i.e., that the complex-balance condition allows us to generalise to non-ideal CRNs the result \eqref{ideal_cb_ness} \cite{Anderson2010ProductFormCRN} for the steady-state distribution of the network, which is generally unique (for details see Refs.  \cite{Schnakenberg1976Network,anderson2015stochastic}).

Namely, we claim that  CRNs for which the complex-balance condition \eqref{complex_bal} holds, 
the steady state of the stochastic dynamics \eqref{CME} with propensity functions \eqref{gen_noneq_rates} reads
\begin{equation}
\label{claim}
\pi_{\rm neq} (\bm{n})=\frac{e^{-\beta [F(\bm{n})+\sum_a \tilde{\mu}_a n_a]}}{Z},
\end{equation}
where the parameters $\tilde{\mu}_a$ depend on the chemostats to which the system is connected and on the reaction constants of the network, but not on $F$. These  parameters can be obtained from the CRN in the ideal and deterministic limit, 
thus significantly simplifying the task of obtaining analytically the steady-state of the system. Note that we reserve $\mu_b$ for the chemical potentials of the chemostats while $\tilde{\mu}_a$ is a parameter that describes how the particle numbers at steady state of the non-chemostatted species depend on the non-equilibrium driving of the system. An additional necessary condition to prove this result is that the function $g_\rho$ must be the same for all reactions, i.e., $g_\rho=g$; the relaxation of this hypothesis will be discussed in Section \ref{disc}. 

The proof follows closely that of Anderson, Craciun and Kurtz \cite{Anderson2010ProductFormCRN}, and  here we only present its main steps---see Appendix \ref{proof_CBdist}  for a full proof. We will substitute the steady-state \eqref{claim} into the dynamical equations, look for  solutions where the probability flux across complexes vanishes, and obtain the complex-balance condition for a network with MAK, \cref{CB_MAK}. We can thus conclue that, if the network modelled deterministically with MAK is complex-balanced at steady state, i.e. \cref{CB_MAK} is satisfied, then \cref{claim} is the steady-state probability distribution of its stochastic non-ideal counterpart. Furthermore, the parameters $\tilde{\mu}_a$ in \cref{claim} can be obtained by solving \cref{CB_MAK}.

By inserting the ansatz \eqref{claim} in \cref{CME} with propensity functions of the form \eqref{gen_noneq_rates} and $g_\rho=g$ for all reactions, at steady state we obtain 

\begin{align}
\sum_\rho  \tilde{k}_\rho \Big \{ g(\bm{n}-\bm{s}^\rho) e^{\beta [ F(\bm{n}) -F(\bm{n}-\bm{s}^\rho)+\sum_a v_a^\rho \tilde{\mu}_a +\sum_b r^\rho_b \mu_b ] } 
+ g(\bm{n}-\bm{r}^\rho) e^{\beta [ F(\bm{n}) -F(\bm{n}-\bm{r}^\rho)  -\sum_a v_a^\rho \tilde{\mu}_a +\sum_b s^\rho_b \mu_b] } &  \Big\} \nn  \\ 
= \sum_\rho \tilde{k}_\rho \Big\{  g(\bm{n}-\bm{r}^\rho) e^{\beta [ F(\bm{n}) -F(\bm{n}-\bm{r}^\rho) +\sum_b r^\rho_b \mu_b] } 
+  g(\bm{n}-\bm{s}^\rho) e^{\beta [ F(\bm{n}) -F(\bm{n}-\bm{s}^\rho)+\sum_b s^\rho_b \mu_b]  } \Big\}&.
\end{align}

The previous equation is satisfied if, for each complex $\bm{z}$, we have 

\begin{align}\label{eq_1}
\sum_{\rho\vert\bm{s}^\rho=\bm{z} }  \tilde{k}_\rho  g(\bm{n}-\bm{s}^\rho)  e^{\beta [ F(\bm{n}) -F(\bm{n}-\bm{s}^\rho) +\sum_a v_a^\rho \tilde{\mu}_a +\sum_b r^\rho_b \mu_b] } +\\ \nn
 \sum_{\rho\vert\bm{r}^\rho=\bm{z} } \tilde{k}_\rho g(\bm{n}-\bm{r}^\rho) e^{\beta [ F(\bm{n}) -F(\bm{n}-\bm{r}^\rho) -\sum_a v_a^\rho \tilde{\mu}_a +\sum_b s^\rho_b \mu_b] }& = \\ \nn
  \sum_{\rho\vert\bm{r}^\rho=\bm{z} } \tilde{k}_\rho g(\bm{n}-\bm{r}^\rho) e^{\beta [ F(\bm{n}) -F(\bm{n}-\bm{r}^\rho)+\sum_b r^\rho_b \mu_b] } +\\ \nn
\sum_{\rho\vert\bm{s}^\rho=\bm{z} } \tilde{k}_\rho g(\bm{n}-\bm{s}^\rho) e^{\beta [ F(\bm{n}) -F(\bm{n}-\bm{s}^\rho)+\sum_b s^\rho_b \mu_b]  }&.
\end{align}
Given that in the previous equation the complex $\bm{z}$ is fixed, it can be simplified and yields 
\begin{align}\label{eq_2}
\sum_{\rho\vert\bm{s}^\rho=\bm{z}}  \tilde{k}_\rho   e^{\beta [ \sum_a (z_a-r_a^\rho) \tilde{\mu}_a +\sum_b r^\rho_b \mu_b]} + \sum_{\rho\vert\bm{r}^\rho=\bm{z}} \tilde{k}_\rho e^{\beta [-\sum_a (s_a^\rho-z_a) \tilde{\mu}_a +\sum_b s^\rho_b \mu_b] } & =\\ \nn
\sum_{\rho\vert\bm{r}^\rho=\bm{z} } \tilde{k}_\rho e^{\beta \sum_b r^\rho_b \mu_b} +\sum_{\rho\vert\bm{s}^\rho=\bm{z} } \tilde{k}_\rho e^{\beta \sum_b s^\rho_b \mu_b }&.
\end{align}
Setting  
\begin{align}\label{eq9a}
c_a^\ast&=\exp[-\beta ( \tilde{\mu}_a+\mu_a^0)], \\
k_{+\rho}&=\tilde{k}_{\rho}\exp \Bigg[\beta \Bigg(\sum_a r_a^\rho \mu_a^0+\sum_b r_b^\rho \mu_b\Bigg)\Bigg] \label{k+}, \\
k_{-\rho}&=\tilde{k}_{\rho}\exp \Bigg[\beta \Bigg(\sum_a s_a^\rho \mu_a^0+\sum_b s_b^\rho \mu_b\Bigg)\Bigg] \label{k-},
\end{align}
\cref{eq_2} can be shown to be equivalent to the complex-balance condition for deterministic CRNs with MAK, \cref{CB_MAK}, with rate constants given by \cref{k+,k-}. These rate constants include the contribution of the standard-state chemical potentials $\mu_a^0$ and the chemostats, as is usually the case in MAK \cite{Esposito2016noneqCRN} (although, without loss of generality, for the purposes of this result, all $\mu^0_a$ can taken to be 0). Hence, a CRN for which the deterministic steady-state is complex balanced allows for a steady state of the form \eqref{claim} for its stochastic and non-ideal version. Solving \cref{CB_MAK} for the steady-state concentrations with MAK and rate constants (\ref{k+}) and (\ref{k-}) yields $c_a^\ast$ and, thus, the parameters $\tilde{\mu}_a$ [via \cref{eq9a}] which appear in the steady-state distribution \eqref{claim}. The exponential relationship between the concentrations $c_a^\ast$ and $\tilde{\mu}_a$ reflects the logarithmic contribution of concentrations in the ideal chemical potential of solutes: $\mu_a=\mu_a^0+1/\beta \log c_a$.

\Cref{claim} shows that the steady-state distribution of a non-ideal complex-balanced CRN has the form of an effective Boltzmann distribution, 
with the standard-state chemical potentials $\mu_a^0$ shifted by $\tilde{\mu}_a$ (typically $\mu_a^0$ would be included within $F$). 
From the physical standpoint, it is interesting to note that in \cref{claim} the free-energetic contribution $F$ and the non-equilibrium term $\sum_a \tilde{\mu}_a n_a$ factor out. 

This result is similar to Theorem 6.6 of Ref. \cite{Anderson2010ProductFormCRN} ---here \cref{Anderson_gen}--- but we have generalised it slightly to include rates of the form (\ref{gen_noneq_rates}), which includes the function $g$ that could be of interest in phase-separated systems as it modulates the rates depending on the environment. 
Moreover, our approach relates both the rates (\ref{gen_noneq_rates}) and the steady-state distribution (\ref{claim}) to thermodynamic quantities, like free energies and chemical potentials.

In what follows, we will  illustrate the result \eqref{claim} with a minimal working example of a complex-balanced CRN, and compare its predictions with numerical simulations. 

\subsection{Example}\label{ex1}

\begin{figure}
\centering
\includegraphics[scale=0.14]{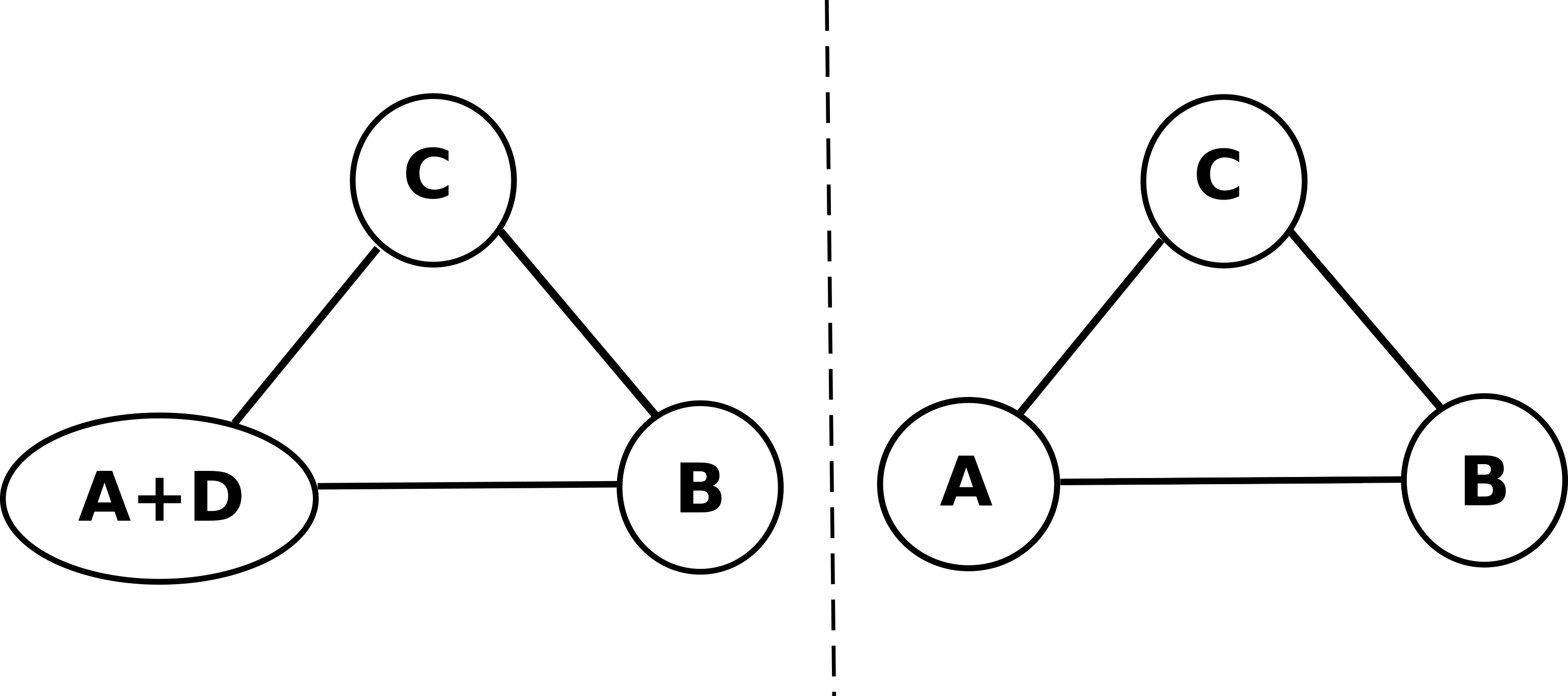}
\caption{Graphical representation of the illustrative CRNs of Sections \ref{ex1} (left) and \ref{ex_Lyap} (right). 
}
\label{CRN_graph}
\end{figure}

Let us consider the following CRN---see Fig. \ref{CRN_graph} for a graphical representation:
\begin{align}
\label{example_SS}
{\rm A}+{\rm D} \leftrightharpoons &\,{\rm B}, \\ \nn
{\rm B} \leftrightharpoons& \, {\rm C},\\ \nn
 {\rm C} \leftrightharpoons &\, {\rm A}+{\rm D}, 
\end{align}
with a free energy taken from a regular-solution theory (where each particle, including the solvent, occupies a finite volume and thus total volume is linked to the total number of particles), see Appendix \ref{virial_rates} for details. 

For the sake of simplicity, we assume that the solvent particle number, $n_\textrm{sol}$, is conserved, and allow the total volume to vary: 
\be
\label{claw1}
N=N_{\rm ABC}+n_{\rm D} +n_\textrm{sol},
\ee 
where the total number of particles  of species $\rm A$, $\rm B$ and $\rm C$,
\be
\label{claw2}
N_{\rm ABC} =n_{\rm A}+n_{\rm B}+n_{\rm C},
\ee 
is kept constant in the CRN defined in (\ref{example_SS}). 

Since the CRN \eqref{example_SS} is complex balanced (which can be checked \textit{a posteriori}), its steady state in the stochastic description and with propensity functions \eqref{gen_noneq_rates} can be obtained from its deterministic dynamical equations \eqref{deterministic_kinetics}.
To achieve this, we write the stoichiometry matrices 
\begin{equation}
\label{matrix_ex}
\bm{R}=
\begin{pmatrix}
1 & 0 & 0 & 1  \\
0 & 1 & 0 & 0 \\
0 & 0 & 1 & 0 
\end{pmatrix},\,
\bm{S}=
\begin{pmatrix}
0 & 1 & 0 & 0  \\
0 & 0 & 1 & 0 \\
1 & 0 & 0 & 1 
\end{pmatrix} ,
\end{equation}
which, together with the reaction constants given by \cref{k+,k-} and the free energy (\ref{Fid}), completely define an ideal CRN. For simplicity, we assume that the standard-state chemical potentials $\mu^0$ take the value $0$ and that $\tilde{k}_\rho=1$ for every reaction $\rho$. Finally, as an example, we take the non-equilibrium contribution of the chemostats to be present only in the reaction C$\leftrightharpoons$A+D, with $\sum_b r_b \mu_b=0$ and $\sum_b s_b \mu_b=\log(5/2) \beta^{-1}$. These considerations, together with MAK [\cref{MAK_det}] and the  dynamics (\ref{deterministic_kinetics}), yield the following set of deterministic and ideal equations for the CRN:
\begin{align}
\label{dyn_ex}
\frac{d n_{\rm A}}{dt}&=n_{\rm B}+\frac{5}{2}\, n_{\rm C}-2 \, n_{\rm A} n_{\rm D}  \\
\frac{d n_{\rm B}}{dt}&=n_{\rm C}+ n_{\rm A} n_{\rm D} -2 \, n_{\rm B}\nonumber \\
\frac{d n_{\rm C}}{dt}&=n_{\rm B}+ n_{\rm A} n_{\rm D} -\frac{7}{2} \, n_{\rm C} .\nonumber 
\end{align}
Note that in the system derived from the matrices (\ref{matrix_ex}) $\textrm{d}_t n_{\rm A}=\textrm{d}_t n_{\rm D}$. For the sake of concreteness, we take as initial conditions 
\be\label{cons1}
N_{\rm ABC}=40
\ee 
and 
\be\label{cons2}
n_{\rm A}-n_{\rm D}=5,
\ee 
the solution of the above system at steady state is $n^\ast_{\rm A}\simeq 8.1$, $n^\ast_{\rm B} \simeq 19.1$, $n^\ast_{\rm C} \simeq 12.7$ and $n^\ast_{\rm D}\simeq 3.1$. It can be checked that this steady-state solution satisfies the complex-balance requirement for MAK, \cref{CB_MAK}. According to \cref{eq9a}, we have the following identity:
$n^\ast_a=e^{-\beta \tilde{\mu}_a}$, which enables us to obtain the values of $\tilde \mu_a$ and  the steady-state probability \eqref{claim}. 

Note that there are two conservation laws, \cref{cons1,cons2}, and four chemical species: hence,  $\pi_{\rm neq} (\bm{n})$ is a distribution with only two independent variables.

In order to evaluate \cref{claim} explicitly,  let us assume that the system has the following regular-solution free energy
\begin{equation}
\label{Free_ex}
\ F(\bm{n})= \frac{1}{\beta} \log \left( \frac{n_{\rm A}!\, n_{\rm B}!\, n_{\rm C}!\, n_{\rm D}!\, n_{\rm sol}!}{N!}\right)  + (n_{\rm A}\mu^0_{\rm A}+n_{\rm C}\mu^0_{\rm C}+n_{\rm D}\mu^0_{\rm D})+\chi \frac{n_{\rm A} n_{\rm C}}{N},
\end{equation}
where the first addend is an entropic term, the second corresponds to the internal energies of the chemical species taken with respect to that of species $\rm B$, and the third to an interaction between species $\rm A$ and $\rm C$. 

\begin{figure*}
\centering
\includegraphics[width=\textwidth]{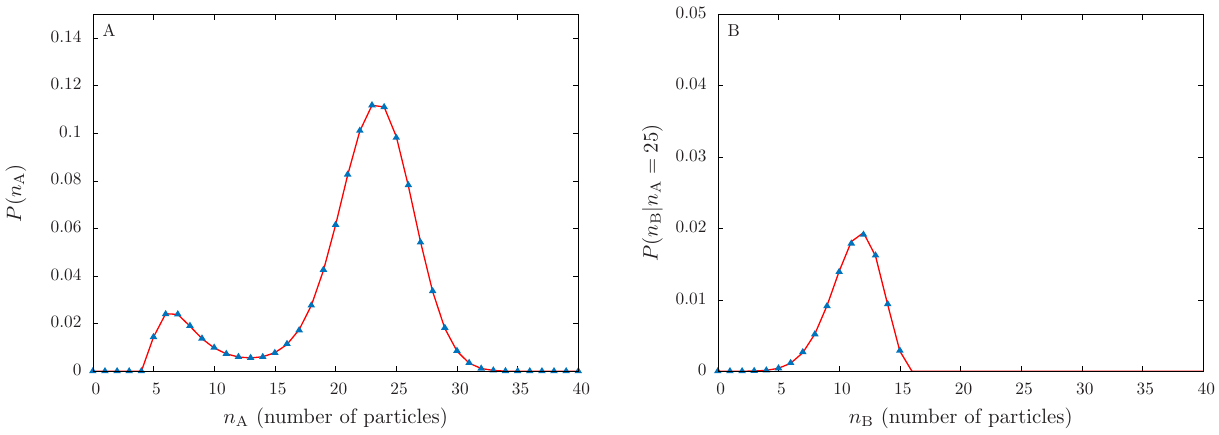}
\caption{Examples of marginal probabilities (A) and conditional probabilities (B)  of the CRN discussed in Section \ref{ex1}. Blue triangles are the probabilities obtained from the Gillespie simulation, and the red lines theoretical predictions from our analytical result \eqref{claim}.}
\label{probs_SS}
\end{figure*}

Setting $\chi=10$ and $\mu^0_{\rm A}=\mu^0_{\rm C}=\mu^0_{\rm D}=\log 2$, we obtain the bimodal steady-state probability depicted in Fig. \ref{probs_SS}, which closely matches the one obtained from a simulation of the same CRN using the Gillespie algorithm \cite{Gillespie1977algorithm}. Simulations were started in parallel from random Poissonian initial conditions satisfying the constraints above,  and the samples were obtained after the simulation relaxed to steady state. Note that, in order to arrive to the set of Eqs. (\ref{dyn_ex}), we assumed all $\mu^0_a=0$, while in the free energy we are giving them a different value. It would have been equivalent to insert these values of $\mu^0_a$ into the system (\ref{dyn_ex}) and omit them in the free energy (\ref{Free_ex}).

\section{Lyapunov function for complex-balanced steady states}\label{sec_lyap}

A Lyapunov function is a function that is minimised by the dynamics of the system and takes the value 0 at steady state. Under fairly general conditions, the logarithm of the steady-state probability distribution in the stochastic CRN is a Lyapunov function of the deterministic one \cite{gang1986lyapounov,ge2017mathematical}. 

While the exact form of the Lyapunov function has been obtained for ideal and complex-balanced CRNs \cite{Horn1972,Anderson2015Lyapunov}, here we demonstrate that for non-ideal, complex-balanced CRNs  the following function decreases with the dynamics
\begin{align}
\label{Lchem}
\mathcal{L}(\bm{c})&=\\ \nn 
-\lim_{V\rightarrow \infty} \frac{1}{V} \log [\pi_{\rm neq}(\bm{n})]&=\\ \nn
\beta\left[\mathfrak{f}(\bm{c})+\sum_a \tilde{\mu}_a c_a\right]+\frac{\log Z}{V}&,
\end{align} 
where the factor $1/V$ has been inserted to maintain the magnitude intensive while $V\rightarrow \infty$.
Our approach  generalises the results of Anderson and Nguyen  \cite{Anderson2019Results} for product-form stationary states of CRNs. 
We will  call the function \eqref{Lchem} a Lyapunov function: This is a slight abuse of terminology, because we will 
only prove that $\mathcal L $ decreases with the dynamics, not  that its value is zero at steady state.

$\mathcal L$ only takes the zero value if
\begin{equation}
\frac{\log Z}{V}= -\beta\left[\mathfrak{f}(\bm{c}^\ast)+\sum_a \tilde{\mu}_a c_a^\ast\right],
\end{equation}
where the asterisk denotes values at steady state. Given that $Z$ is a normalisation factor for the stochastic complex-balanced CRN at steady state, see \cref{claim}, it reads
\begin{equation}
Z=\sum_{\bm{n}} e^{-\beta [F(\bm{n})+\sum_a \tilde{\mu}_a n_a]},
\end{equation}
which, for a large (many particles) deterministic CRNs, can be evaluated using the saddle-point approximation, where the sum is evaluated at the minimum value of the argument of the exponential. If the deterministic system is monostable, then the argument of the exponential has a single local minimum. Therefore, for monostable CRNs, this approximation will yield the correct value and the Lyapunov function \cref{Lchem} will take the value $0$ at steady state. However, care must be taken when handling multistable CRNs in this way, which is why, in order to avoid this complexities, we will not prove that \cref{Lchem} takes the value $0$ at steady state in general.  Nevertheless, the fact that this function decreases with the dynamics is sufficient for our purposes.
\vspace{0.5cm}

In what follows  we sketch the proof that, for complex-balanced non-ideal CRNs, the Lyapunov function \eqref{Lchem} is a decreasing function of  time---for a full step-by-step proof, see Appendix \ref{proof_minimization}.

Given that the normalisation factor $Z$ does not depend on time but only on the non-equilibrium steady state, the time derivative of $\mathcal L$ is
\begin{align}\nn
\frac{d  \mathcal{L}}{dt}=&\\ \nn
\sum_{k} \frac{\partial \mathcal{L}}{\partial c_{k}} \frac{\partial c_{k}}{\partial t}  =& \\ 
\beta \sum_{k} (\mu_{k}+\tilde{\mu}_{k}) \left\{  \sum_\rho v_{k}^\rho k_\rho g(\bm{c}) \left[e^{\beta( \sum_a r_a^\rho \mu_a+ \sum_b r_b^\rho \mu_b)} -e^{\beta (\sum_a s_a^\rho \mu_a+\sum_b s_b^\rho \mu_b)}\right] \right\},
\end{align}
where in the second line we used \cref{deterministic_kinetics,gen_noneq_rates_det}, together with the assumption $g_\rho=g$. After adding and subtracting terms of the form $\sum_a r_a^\rho \tilde{\mu}_a$ in the exponentials (of the form $\sum_a s_a^\rho \tilde{\mu}_a$ for the second exponential), we repeatedly apply  the inequality $e^x(y-x)\le e^y-e^x$  to the sums of chemical potentials, and obtain
\begin{align}
\frac{d  \mathcal{L}}{dt} \le &\sum_\rho  k_\rho g(\bm{c}) e^{\beta (\sum_b r_b^\rho \mu_b- \sum_a r_a^\rho \tilde{\mu}_a)} \left[ e^{ \beta \sum_a (\mu_a+\tilde{\mu}_a)s_a^\rho}- e^{\beta \sum_a r_a^\rho (\mu_a+\tilde{\mu}_a)}\right]+  \\ \nn
 & \sum_\rho k_\rho g(\bm{c}) e^{\beta (\sum_b s_b^\rho \mu_b- \sum_a s_a^\rho \tilde{\mu}_a) }  \left[  e^{\beta  \sum_a(\mu_a+\tilde{\mu}_a)r_a^\rho}- e^{\beta \sum_a s_a^\rho (\mu_a+\tilde{\mu}_a)}\right].
\end{align}

The expression  in the right-hand side (RHS) above can be split in terms of the different complexes in the system:
\begin{align}\label{eq3}
\frac{d  \mathcal{L}}{dt}\le  & \\ \nn
  \sum_{\bm{z} } g(\bm{c}) \Bigg \lbrace & \sum_{\rho\vert\bm{s}^\rho=\bm{z}} k_\rho  e^{\beta [\sum_b r_b^\rho \mu_b-\sum_a r_a^\rho \tilde{\mu}_a+ \sum_a (\mu_a+\tilde{\mu}_a)s_a^\rho]}-
  \sum_{\rho\vert\bm{r}^\rho=\bm{z}} k_\rho  e^{\beta [\sum_b r_b^\rho \mu_b-\sum_a r_a^\rho \tilde{\mu}_a+ \sum_a r_a^\rho (\mu_a+\tilde{\mu}_a)]}+ \\  \nn &
 \sum_{\rho\vert\bm{r}^\rho=\bm{z}}  k_\rho e^{\beta [\sum_b s_b^\rho \mu_b-\sum_a s_a^\rho \tilde{\mu}_a +   \sum_a(\mu_a+\tilde{\mu}_a)r_a^\rho]}- 
\sum_{\rho\vert\bm{s}^\rho=\bm{z}}  k_\rho e^{\beta [\sum_b s_b^\rho \mu_b- \sum_a s_a^\rho \tilde{\mu}_a+ \sum_a s_a^\rho (\mu_a+\tilde{\mu}_a)]}\Bigg \rbrace.
\end{align}

For a complex-balanced system, it can be shown that the expression in curly brackets in \cref{eq3} vanishes for each complex $\bm{z}$ independently, as a consequence of the complex-balance condition for MAK systems, \cref{CB_MAK}. Then
\begin{equation}\label{der_L}
\frac{d  \mathcal{L}}{dt} \le 0,
\end{equation}
and $\mathcal{L}$ decreases, or remains unchanged, along a trajectory.

We conclude that, unlike in classical equilibrium systems, here it is not the $F$ that is minimised by the dynamics, but a free energy \eqref{Lchem} where the standard chemical potentials $\mu_a^0$ are shifted by $\tilde{\mu}_a$. 
This shift, which is entirely due to the non-equilibrium contribution of the chemostats, enables the system to present non-vanishing chemical-reaction net flows between species at steady-state, which, in the following, we will call chemical currents.

\subsection{Spatially heterogeneous systems}\label{heterogeneous}

In order to describe phase-separating systems, in what follows we will incorporate in our framework spatial inhomogeneities. In the deterministic description, concentrations are now a function of space, $c_a(\bm{x})$ within a volume $\Omega$, and the free energy is a functional of these concentrations, $F[\bm{c}]$.

The time derivative of the concentrations is given by the following reaction-diffusion (RD) equation
\begin{equation}
\label{RDeq}
\frac{\partial c_a(\bm{x})}{\partial t}=-\bm{\nabla} \cdot \bm{J}_a+ \sum_\rho v_a^\rho [J_{+\rho}(\bm{x})-J_{-\rho}(\bm{x})],
\end{equation}
where the time dependence of $c$ is omitted, the first term in the RHS of the equation represents diffusion, and the second one the chemical reactions. As in the 
linear irreversible thermodynamics framework \cite{groot&mazur}, the driving force of the diffusion current $ \bm{J}_a$  is the gradient of chemical potentials, $\bm{\nabla} \mu_a$: the diffusive currents then read 
\be
\bm{J}_a=-\sum_k \mathcal{M}_{ak} \bm{\nabla} \mu_k,
\ee 
where $ \mathcal{M}_{ak}$ is the mobility matrix. We assume no-flux boundary conditions 
\be\label{no_flux}
\bm{J}_a\vert_{\bm{x} \in \partial \Omega}=0
\ee
for the non-chemostatted species, where $\partial \Omega$ denotes the boundaries of the volume $\Omega$.\\

We now consider a generalisation of the  Lyapunov function \eqref{Lchem} to inhomogeneous systems. In the following, we will show that the dynamics (\ref{RDeq}) for complex-balanced networks minimise the Lyapunov functional 
\be
\label{RDLyapunov}
L = \beta \int_{\Omega}  {\rm d} \bm{x} \mathcal{L}(\bm{c}(\mathbf{x})) =   \beta F[\mathbf{c}]+ \beta \int_{\Omega}  {\rm d} \bm{x}  \sum_a \tilde{\mu}_a c_a(\mathbf{x})
\ee
where $F[\mathbf{c}]$ is the free energy of the system, which depends on the concentration profile through
\be
F[\bm{c}] = \int_{\Omega} {\rm d} \bm{x} \, \mathfrak{f}(\bm{c}).
\ee 

The time derivative of the Lyapunov functional (\ref{RDLyapunov}) yields
\begin{align}\label{dLdt}
\frac{d L}{dt} & = \\ \nn
\int_\Omega\frac{\partial \mathcal{L}}{\partial t}{\rm d} \bm{x}&=\\ \nn
 \sum_a \int_\Omega {\rm d} \bm{x} \, \frac{\partial \mathcal{L}}{\partial c_a({\bm x})} \frac{\partial c_a({\bm x})}{\partial t}&=\\\nn
 \sum_a \int_\Omega {\rm d} \bm{x} \, \beta[\mu_a(\bm{x}) + \tilde{\mu}_a] \left[ {\bm \nabla} \cdot \bm{J}_a + \sum_\rho v_a^\rho (J_{+\rho}-J_{-\rho})\right]&,
\end{align}
where $\mu_a(\bm{x})=\delta F[\bm{c}(\bm{x})]/\delta c_a(\bm{x})$ is the local chemical potential.

By applying the results of Section \ref{sec_lyap}, \cref{der_L}, at every spatial point $\bm{x}$, we obtain that the  second term in the square brackets of the RHS of \cref{dLdt} is negative or zero. Therefore, to prove that $dL/dt \leq 0 $ it is sufficient to show that the first term in the square brackets of  the RHS is also negative or zero. In this regard, we note that 
 \begin{align}
 &\sum_a \int_\Omega {\rm d} \bm{x} \, [\mu_a (\bm{x}) + \tilde{\mu}_a] {\bm \nabla} \cdot \bm{J}_a \nn \\ & = \sum_a \int_\Omega {\rm d} \bm{x}  \lbrace {\bm \nabla} \cdot [(\mu_a (\bm{x} ) + \tilde{\mu}_a) \bm{J}_a ] -[{\bm \nabla} (\mu_a (\bm{x}) + \tilde{\mu}_a)] \cdot \bm{J}_a \rbrace.
\end{align}
The first term in the RHS of the last equality vanishes due to the divergence theorem and Neumann  boundary conditions \eqref{no_flux}. By observing that $\tilde{\mu}$ does not depend on space, the addend containing $\tilde{\mu}$ in the second term vanishes ($\nabla \tilde{\mu}_a=0$). Finally, if the Onsager reciprocal relations for the mobility matrix $ \mathcal{M}_{ak}$ hold \cite{onsager1931reciprocal}, then the addend containing $\mu_a({\bm x})$ in the second term is necessarily positive, because it represents the entropy production of a diffusion process \cite{prigogineThermodynamics,groot&mazur}.
The Onsager reciprocity relations ensure that a system relaxes to equilibrium in the absence of external work. Thus, the condition that the Onsager relations hold is not a limitation of the result but a consequence of thermodynamical consistency.

Combining the results above, we obtain that
\begin{equation}
\frac{dL}{dt} \le 0.
\end{equation}
It follows that, for a non-ideal, complex-balanced system, $L$ decreases, which we can now use to obtain useful information about the steady state, along the lines of the free energy minimisation for systems at thermodynamic equilibrium.

Therefore, for a complex-balanced system, we can minimise $L$ (subject to constraints in particle numbers) in order to obtain the concentration profile at steady state. This minimisation results in constraints for the steady-state profile of the form
\begin{equation}
\label{const_SS}
\frac{\delta L}{\delta c_a(\mathbf{x})}= \beta (\mu_a (\mathbf{x})+\tilde{\mu}_a) -\lambda =0,
\end{equation}
where $\lambda$ is a Lagrange multiplier that enforces the particle-conservation constraint---for further details see Section \ref{chem_reac_mix}. \Cref{const_SS} implies that at steady-state in a complex-balanced solution there cannot be any diffusive currents, since the chemical potential is constant throughout space and the force driving diffusion currents is $\nabla \mu_a({\bm x})$. Nevertheless, chemical currents can exist at steady state, as noted in the previous section, and the concentration profile may not be homogeneous. This is a major consequence of the present work.

\subsubsection{Example}\label{ex_Lyap}

Let us consider the following CRN (see Fig. \ref{CRN_graph} for a graphical representation):
\begin{align}
\label{example_Lyap}
{\rm A} \leftrightharpoons {\rm B}\, \\\nn
 {\rm B} \leftrightharpoons {\rm C}, \\ \nn
 {\rm C} \leftrightharpoons {\rm A}, 
\end{align}
with a free energy taken from a regular solution theory, as before. We will first assume that the system is homogeneous and later we will analyse the full reaction-diffusion system. 

For simplicity, we assume that the system is driven out of equilibrium solely by imposing a non-equilibrium chemical potential difference in the transition from $\rm C \leftrightharpoons \rm A$, with $\sum_b s_b \mu_b=-\Delta \beta^{-1}$ and $\sum_b r_b \mu_b=0$. 

We take all the reaction constants $k$ equal to each other, and note that the network is necessarily complex-balanced, as all chemical reactions are unimolecular (in unimolecular networks each of the species is a complex, hence the steady-state condition is equivalent to the complex-balance condition, if $g_\rho=g$ for every reaction $\rho$).

Proceeding along the lines of Section \ref{ex1},
 in the stochastic description the steady-state of the CRN \eqref{example_Lyap} with propensity functions \eqref{gen_noneq_rates} can be obtained from the following ideal and deterministic rate equations:
\begin{align}\label{eq4}
\frac{d n_{\rm A}}{dt}&=n_{\rm B}+e^{-\Delta}n_{\rm C}-2 n_{\rm A} \nonumber \\
\frac{d n_{\rm B}}{dt}&=n_{\rm C}+ n_{\rm A}  -2n_{\rm B}\nonumber \\
\frac{d n_{\rm C}}{dt}&=n_{\rm B}+ n_{\rm A}  -(1+e^{-\Delta})n_{\rm C} .\nonumber 
\end{align}
The solution of the above system at steady-state is 
\begin{align}
n_{\rm A}&=\frac{2 n_{\rm C}}{3} \left(\frac{1}{2}+e^{-\Delta}\right) , \nn\\ 
n_{\rm B}&=n_{\rm C} \left[ \left(\frac{1}{2}+e^{-\Delta}\right)\frac{1}{3}+\frac{1}{2}\right] \nonumber .
\end{align}
We obtain the values of $\tilde{\mu}_a$ by identifying $n_a$ with $e^{-\beta \tilde{\mu}_a}$ [as in \cref{eq9a}]. Noting that we can express such potentials with respect to that of species $\rm C$, we obtain the Lyapunov function of the system
\begin{equation}
\label{eqexampleLyap}
\mathcal{L}(\bm{c})=\beta \mathfrak{f}(\bm{c})-c_{\rm A} \log \frac{1+2 e^{-\Delta}}{3}-c_{\rm B} \log \left( \frac{1}{2} + \frac{\frac{1}{2}+e^{-\Delta}}{3}\right)-\frac{\log Z}{V},
\end{equation}
where a detailed expression of the normalisation constant $Z$ is not essential here, because $Z$ is constant along the dynamics, and it does not alter the location of the minima of $\mathcal{L}$ in the space of concentrations $\bm{c}$.\\

 Unlike above, we will now describe the amount of species with reference to the fraction of volume they occupy at each point of space $\phi(\mathbf{x})$. Then, 
 \begin{equation}
 \label{vol_fracs}
 \sum_a \phi_a(\mathbf{x}) +\phi_\textrm{solv} (\mathbf{x})=1,
 \end{equation}
  where $\phi_\textrm{solv} (\mathbf{x})$ is the volume fraction of the solvent and the sum runs over solutes only. \Cref{vol_fracs} states that the solution is incompressible and, thus, $\phi_\textrm{solv} (\mathbf{x})=1-\sum_a \phi_a(\mathbf{x}).$ The reason for using volume fractions instead of concentrations is threefold: It is the convention normally used in phase separation studies and regular solution models, it enforces incompressibility (which is the case in most liquids) and is dimensionless. For simplicity we will assume that the molecular volumes of every species is the same, so that $\phi_a(\mathbf{x})$ is proportional to $c_a(\mathbf{x})$.

Hence, the following regular solution free-energy density can describe spatial inhomogeneities in an incompressible solution:
\begin{align}
\beta \mathfrak{f}({\bm c})=&\sum_a \phi_a \log \phi_a+ \Bigg(1-\sum_a \phi_a \Bigg)\log \Bigg(1- \sum_a \phi_a \Bigg)\nonumber \\ \label{eq5} &  + \chi_{\rm AA} \phi_{\rm A}^2+\chi_{\rm AB} \phi_{\rm A} \phi_{\rm B}+\sum_{a,k}\kappa_{a,k}(\bm{\nabla} \phi_a) \cdot \bm{\nabla} \phi_{k},
\end{align}
 where the first two terms in the RHS are entropic terms and the following two represent the interactions between the solutes. The last addend represents the free-energetic cost of spatial inhomogeneities in the concentration profiles, and  is known as Cahn-Hilliard term \cite{cahn1958}.

Assuming the system is one-dimensional, the resulting Lyapunov functional for the RD system is
\be
\label{RDLyapunov_ex}
L = \beta \int {\rm d} x \Bigg[ \mathfrak{f}(\bm{\phi})+ \sum_a \tilde{\mu}_a \phi_a(x) \Bigg],
\ee
where $\tilde{\mu}_a$ are the ones obtained for the homogeneous system and do not depend on the coordinate $x$.
 
\begin{figure*}[h!]
\centering
\includegraphics[width=0.76\textwidth]{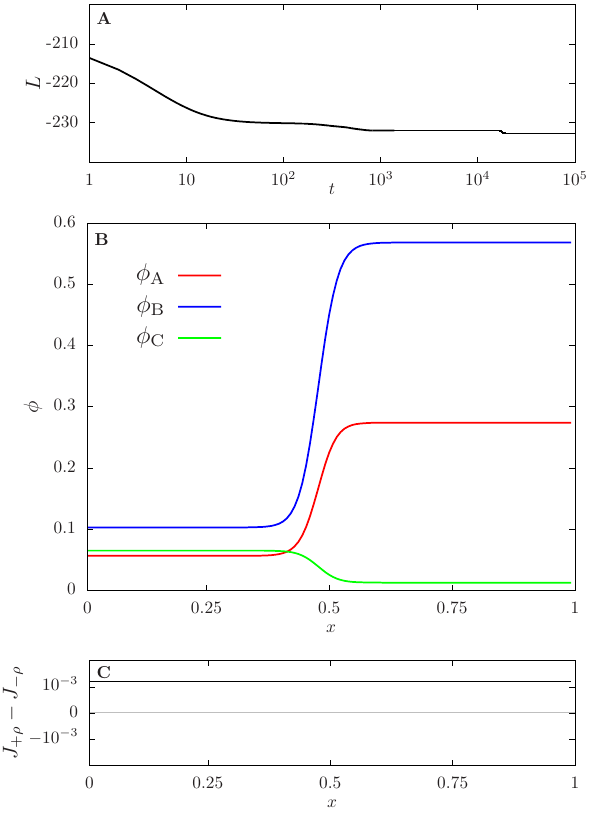}
\caption{Numerical results for the CRN (\ref{example_Lyap}) obtained by perturbing a homogeneous solution and integrating forward in time the reaction-diffusion equations (\ref{RDeq}) until a steady state is reached. The system is considered one-dimensional, and its normalised spatial coordinate $x$ lies between zero and one. A) Evolution of the Lyapunov functional as a function of time in logarithmic scale. Here, the Lyapunov functional does not include the constant  term $\log Z$ in \cref{Lchem}. 
B) Steady-state profile of the system, displaying phase coexistence. The volume fractions $\phi$ of species $\rm A$, $\rm B$ and $\rm C$ are plotted as functions of  $x$. C) Net reaction flux $J_{+\rho}-J_{-\rho}$ (black line, arbitrary units) at steady state, as a function of $x$.}
\label{Lyap_ex}
\end{figure*}

 We set $\chi_{\rm AA}=-2$,  $\chi_{\rm AB}=-7$, $\Delta=-2$, $\kappa_{\rm A,A}=\kappa_{\rm B,B}=5$, $\kappa_{\rm A,B}=1$ and any other Cahn-Hilliard coefficient equal to 0. With this parameter set, the reaction-diffusion system exhibits phase separation at steady state (see Fig. \ref{Lyap_ex}). By entering this free energy in  the RD equations \eqref{RDeq} and assuming no state dependency of the reaction constants $k_\rho$, we obtain a set of equations 
which describes the dynamics of the system. Figure \ref{Lyap_ex} shows that the Lyapunov functional \eqref{RDLyapunov_ex} is minimised by the dynamics, and that the non-equilibrium steady state is characterised by  phase coexistence.

Finally, in Fig \ref{Lyap_ex} C, the net reaction flux $J_{+\rho}-J_{-\rho}$ at steady state as a function of the spatial coordinate is depicted. This net reaction flux is constant in space and, given the topology of the CRN (\ref{example_Lyap}),  is equal for all reactions $ \rho$. The fact that the net reaction flux is independent of the spatial coordinate despite the varying concentrations (see Fig. \ref{Lyap_ex} B) is a result of chemical reaction fluxes being driven by the chemical potential, which, as argued above, is constant---see \cref{const_SS}. Note that this is also a consequence of having dropped the dependency of the reaction constants on the environment via a function $g(\bm{c})$. If all reaction constants were subject to this modulation (which has to be the same for every reaction for our results to hold), then the reaction rates at steady state could be space-dependent but the chemical potential would still be constant.

\subsection{Phase Diagram of a chemically reactive mixture}\label{chem_reac_mix}

Since the  Lyapunov functional for complex-balanced systems discussed in Section \ref{heterogeneous} is minimised by the dynamics, it carries plenty of information on the steady state. 

Along the lines of phase separation for equilibrium systems, the steady state can be obtained by minimising $L[\bm{c}]$ with respect to $\bm c$, subject to certain constraints, e.g., particle conservation.  The concentration profiles which realise the absolute minimum of $L[\bm{c}]$ may be either spatially uniform, or depend on space, according to the system parameters. 
On a qualitative level, the phenomenology of a complex-balanced system does not change much with respect to that of a non-ideal solution at equilibrium, but the non-equilibrium terms may alter the phase diagram, thus tweaking the onset of phase separation. 

To illustrate this point, in  this Section we consider a non-ideal solution with the CRN \eqref{example_Lyap} in the deterministic description, and obtain its phase diagram. Therefore, we minimise the Lyapunov functional \eqref{RDLyapunov_ex} of Section \ref{ex_Lyap}, with the particle-conservation constraint 
\be
\phi_N=\frac{1}{\vert\Omega\vert}\int_\Omega {\rm d} \bm{x}[\phi_{\rm A}(\bm{x})+\phi_{\rm B}(\bm{x})+\phi_{\rm C}(\bm{x})],
\ee
where $\phi_a(\bm{x})$ is the volume fraction of species $a$, and the constant $\phi_N$ fixes the total volume fraction of the solutes. Then, the function that needs to be minimised is the Lagrangian
\be
\label{Lyap_functional}
\Lambda=
\int_\Omega {\rm d} \bm{x} \, \mathcal{L}(\bm{\phi}) - \lambda \left\{ \vert\Omega\vert \, \phi_N  - \int_\Omega {\rm d} \bm{x}\,[\phi_{\rm A}(\bm{x})+\phi_{\rm B}(\bm{x})+\phi_{\rm C}(\bm{x})]\right\},
\ee
where $\lambda$ is the Lagrange multiplier associated with the conservation of solutes, and $\vert\Omega\vert$ the volume of $\Omega$.  

For the sake of simplicity, we take the typical lengthscale of $\Omega $ to be  large with respect to inter-species interfaces: as a result, the volume fractions $\bm \phi(\bm{x})$ can be approximated by piecewise constant functions. If the system phase separates, we assume that only two homogeneous, distinct phases, which we denote by `$1$' and `$2$', will appear.  Within this assumption, the Lagrangian \eqref{Lyap_functional} reads

\begin{equation} 
\label{Lagrangian2}
\Lambda=\vert\Omega^1\vert\mathcal{L}(\bm{\phi}^1)+\vert\Omega_2\vert\mathcal{L}(\bm{\phi}^2)- \lambda [\vert\Omega\vert \phi_N -\vert\Omega^1\vert(\phi^1_{\rm A}+\phi^1_{\rm B}+\phi^1_{\rm C})-\vert\Omega_2\vert(\phi^2_{\rm A}+\phi^2_{\rm B}+\phi^2_{\rm C})],
\end{equation}
where $\Omega_1$ and $\Omega_2$ stand for the volumes  phases $1$ and $2$, respectively, with $\vert\Omega_1\vert+\vert\Omega_2\vert = \vert\Omega\vert$, and we consider the following free-energy density
\begin{equation}
\label{free_phiA}
\beta \mathfrak{f}(\bm{\phi})=\sum_a \phi_a \log \phi_a + \Bigg(1-\sum_a \phi_a \Bigg) \log \Bigg(1-\sum_a \phi_a \Bigg)- \chi \phi_{\rm A}^2,
\end{equation}
where all species except A are considered non-interacting and A interacts with itself ($\chi>0$ implies an effective attraction between A particles).

\begin{figure}[t]
\centering
  \includegraphics[width=\textwidth]{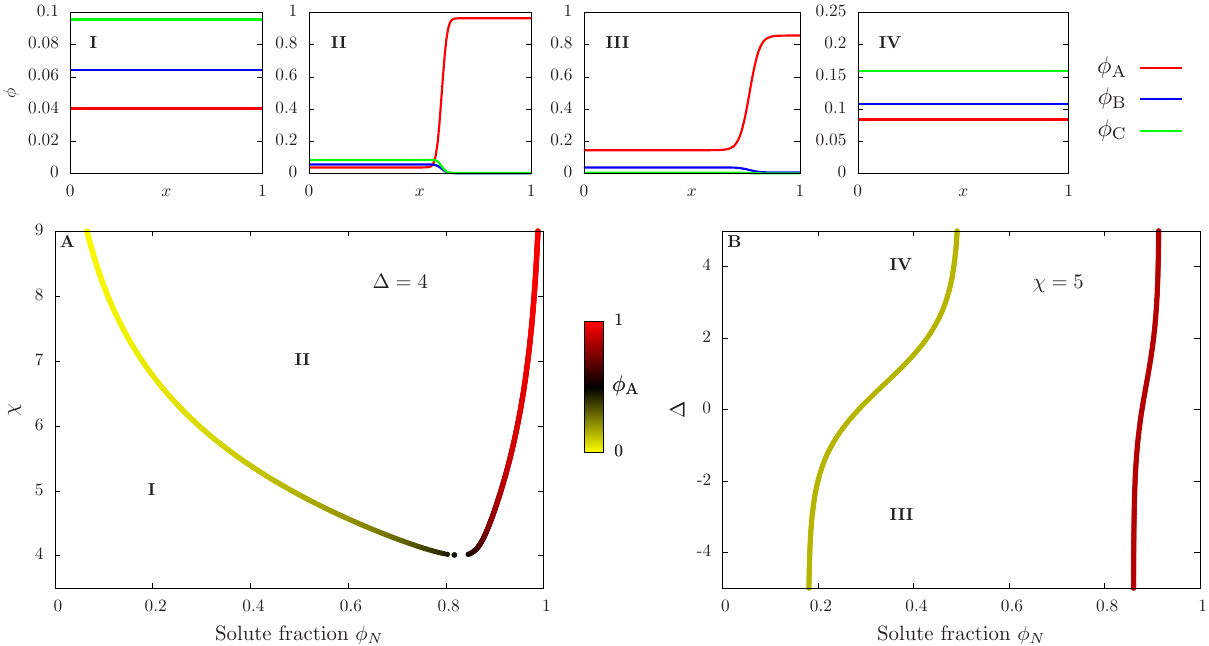}
\caption{
Top: Steady-state concentration profiles of the example discussed in Section \ref{chem_reac_mix}, obtained by numerical integration of the RD equations [\cref{RDeq}]. The concentrations of each species are plotted as functions of the one-dimensional space coordinate $x$, from $x=0$ to $x=1$. Each panel has a number (roman numerals) associated and corresponds to a different parameter configuration, which is specified by the exact position of the roman numerals in the phase diagram below. The leftmost and rightmost panel correspond to a spatially uniform steady state, while in the two middle panels phase separation occurs.  
Bottom: Phase diagrams as functions of the total solute fraction $\phi_N$ and interaction parameter $\chi$ (left) or non-equilibrium driving $\Delta$ (right), obtained by minimising the Lagrangian $\Lambda$, see Appendix \ref{minimization}. The color code represents the amount of $A$ particles along the phase-coexistence lines. While in the top panels the width of the inter-phase interface is finite to ease numerical integration, in the lower panels such width has been assumed to vanish.}
\label{phase_diagram}
\end{figure}

The minimisation of $\Lambda$ yields the phase diagram in Fig. \ref{phase_diagram}, see Appendix \ref{minimization} for details. Phase separation occurs in regions 
$\textbf{II}$ and $\textbf{III}$ of the phase diagram, as shown in the concentration profiles displayed in the insets. 

From the form of the free energy \eqref{free_phiA}, we can see that is the species $\rm A$ that drives phase separation, since for $\chi >0$ the free energy will favour segregating A from the rest of the solution. Thus, whether the steady-state displays one phase or a coexistence of phases also depends on the value of the non-equilibrium chemical potential difference $\Delta$, which can alter the concentration of A at steady state and, hence, modulate phase separation, as can be seen in Fig \ref{phase_diagram} \textbf{B}.

\section{Discussion}\label{disc}

In this work, we have shown that for a chemically reactive non-ideal solution we can obtain results for complex-balanced networks analogous to those for ideal  solutions, provided that the system is modelled in a thermodynamically consistent way. This implies that the rates of the chemical reactions incorporate the interactions between the species in the system and, therefore, mass-action kinetics (MAK) no longer holds. By generalising MAK to a non-ideal solution, we obtained the steady-state probability distribution for a stochastic complex-balanced CRN and the Lyapunov function of its deterministic counterpart, which determines the phase diagram of the system.

Our results are of particular importance for non-equilibrium phase-separating systems. By combining previous results from the mathematical theory of CRNs \citep{Anderson2010ProductFormCRN,Anderson2019Results} and concepts of non-equilibrium thermodynamics \cite{onsager1931reciprocal,prigogineThermodynamics},  we found that the resulting complex-balanced RD system cannot sustain diffusion currents at steady state, see \cref{const_SS}. Since, in many cases, diffusion currents are required for pattern formation in reaction-diffusion systems, breaking complex balance is a necessary condition to obtain such patterned  steady states, at least when interactions are modelled in a thermodynamically consistent way unlike, e.g., those in Refs.  \cite{FanLee2018CRCPhase,Li2020noneq}. 
In this regard, complex balance can be broken in two ways: First, by choosing  a suitable network topology that allows for a steady state which is not complex-balanced, as in Ref. \cite{LefeverCarati1997}. Second, in a system where different phases coexist,  by allowing the reaction rates to depend differently on local environment:  For example, in Ref. \cite{Zwicker2021controlling} a patterned steady-state is produced by allowing one (and only one) of the reaction constants to depend on the concentration of an enzyme which localises in one of the phases. Mathematically, this violates one of the necessary conditions for our results to hold, namely $g_\rho=g$ (see Section \ref{sec_claim}), thus allowing for more general steady states.

In biological cells, phase separation has been hypothesised to perform many functions, such as, accelerating biochemical reactions within the condensate irrespective of the rate of the reaction in the dilute phase \cite{Rosen2021Functions,Weis2020Phasing}. The present work implies that, in order to control chemical reactions  in each of the phases independently (at steady state) breaking complex balance is necessary, by virtue of \cref{const_SS}. Indeed, in a complex-balanced system, the chemical potential of every species is constant throughout space. Then, given that the force driving the chemical reactions are the chemical potentials, the reaction rates in both phases are related, making it impossible to regulate the rates of chemical reactions in each phase in a fully independent way, and suggesting that breaking complex balance in one of the two ways outlined above is crucial for such control.


Overall, complex balance is known to be a key feature of CRNs which determines not only their behavior \cite{Anderson2010ProductFormCRN, Feinberg1995} but also their thermodynamic properties \cite{Polettini2015Deficiency}. In this analysis, we further stress the connection between the characteristics of the reaction network and the thermodynamically consistent structure of the physical system, in an effort to generalise results from ideal CRNs, and explore non-equilibrium dynamics of complex-balanced networks. However, little is known about non-complex balanced systems and, given our results, further research regarding the behaviour of this type of networks out of thermodynamic equilibrium would be of the utmost importance, both from the physical \cite{Mehta2016} and biological \cite{Weis2020Phasing} point of view.


\textbf{Acknowledgments.}
We thank F. Brochard-Wyart, A. S. Vishen, P. Sens,  U. Gerland, J.-F. Joanny, D. Lacoste, J. Palmeri, A. \v{S}ari\'c for valuable conversations and suggestions.

\section*{Declarations}


\begin{itemize}
\item Funding: This study is supported by a Agence nationale de la recherche (ANR) grant ANR-17-CE11-0004.
\item Conflict of interest/Competing interests: The authors declare that they have no conflict of interest nor competing interests.
\item Authors' contributions: A.M.M. conceived the study and wrote the paper. M.C. contributed to the discussions and revisions. 
\end{itemize}

\noindent

\bigskip





\begin{appendices}
\section{Form of the propensity functions for a regular-solution theory}\label{virial_rates}

We now consider a model of a solution based on a lattice where each chemical species (including the solvent) occupies one lattice site, thus neglecting differences in molecular volumes. 

In a lattice with $\mathcal{N}$  sites (note that the number of sites is proportional to the volume) occupied by $N$ different species, with $\sum_{a=1}^{N} n_a=\mathcal{N}$, the configurational entropy is given by
\begin{equation}
\label{config_entropy}
S=k_{\rm B}  \log \frac{\mathcal{N}!}{\prod_a n_a!},
\end{equation}
where the argument of the logarithm is the number of microstates. The internal free energy of each species $a$ is given by the standard-state chemical potential $\mu_a^0$. 

We incorporate in this regular-solution model interactions among neighbouring sites, whose energy (in the mean-field approximation) reads
\begin{equation}
\sum_{a, k}  \frac{\chi_{a k} }{2 \mathcal{N}} \left[n_a n_k (1-\delta_{a, k})+n_a(n_a - 1)\delta_{a, k}\right]=\sum_{a, k}  \frac{\chi_{a k} }{2 \mathcal{N}} \left(n_a n_k -n_a \delta_{a, k}\right),
\end{equation}
where $\chi_{a k}$ represents the interaction energy between species $a$ and $k$, and it can also be interpreted as the matrix of virial coefficients.

Taken into account the previous considerations, the free energy for a homogeneous mixture of chemical species in the regular-solution model reads
\begin{align}
\label{free_virial}
F&=U-TS\\ \nn
&=k_{\rm B} T\left[\sum_a \log (n_a !) -\log (\mathcal{N}!)\right] 
+ \sum_a \mu_a^0 n_{a} + \sum_{a, k}  \frac{\chi_{a k} }{2 \mathcal{N}} \left( n_a n_k -n_a \delta_{a, k}\right),
\end{align}
where the first two terms in the last line represent the ideal free energy (see \cref{Fid} in the main text), while the last term is exclusively due to interactions between solutes.

With this expression of the free energy we can now derive an expression for the propensity functions \eqref{gen_rate+} and \eqref{gen_rate-}. The forward (or backward) rates are a function of the free energy difference of the complex:
 \begin{align}
F(\bm{n})-F(\bm{n}-\bm{r}^{\rho})= 
\Delta F_{\textrm{id}}+\sum_a r^{\rho}_a \sum_k \frac{\chi_{a k} }{ \mathcal{N}} n_k-\sum_{a,k} \frac{\chi_{a k} }{2 \mathcal{N}}r^\rho_k r^\rho_a+\sum_{a}  \frac{\chi_{a a} }{2 \mathcal{N}} r^{\rho}_a  \label{DF_regular}
\end{align} 
where $\Delta F_{\textrm{id}}$ is the ideal part of the free energy difference of the complex,  given by \cref{DF_mass_action} in the main text.
In the RHS of \cref{DF_regular}, only the first two terms are non-vanishing as we approach the thermodynamic limit ($\mathcal{N} \rightarrow \infty$, while keeping $n_a/\mathcal{N}$ fixed): hence, for large systems, the rest of the interacting terms are negligible. However, for a unimolecular reaction, since the free energy difference takes a particularly simple form, we have that
\begin{align}
f_{+\rho}(\bm{n})=&k_\rho e^{\beta[F(\bm{n})-F(\bm{n}-\bm{r}^{\rho})]}\\ \nn
=&k_\rho \frac{n_a}{\mathcal{N}} e^{\beta\left(\mu_a^0+\sum_k \frac{\chi_{a \, k} }{ N} n_k\right)},
\end{align}
where $a$ is the reactant of the reaction $+\rho$.

In the thermodynamic limit, \cref{gen_rate+,gen_rate-,DF_regular}  imply that the deterministic rates can be written as 
\begin{equation}
\label{nonid_rate}
J_{+\rho}=k_{+\rho} \prod_a c_a^{r^{\rho}_a} \exp \left( \sum_{a,k} r^{\rho}_a \chi_{a k} c_k \right),
\end{equation}
where the particle numbers have been replaced with concentrations (an additional logarithmic factor has been absorbed into the rate constant $k_{+\rho}$, as explained in the main text, Section \ref{det_dynamics}) and the part of the chemical potential representing the internal energy has also been absorbed in the rate constant $k_{+\rho}$.
Setting
\begin{equation}
\label{virial_mu}
\mu_a= \log c_a  + \mu^0_a+\sum_k \chi_{a k} c_k,
\end{equation}
the rates \eqref{nonid_rate} match the general expression given in the main text, \cref{det_rates}.

\section{Proof of the complex-balanced distribution} \label{proof_CBdist}

In this Section we present the full proof of the result \eqref{claim}.

At steady state, the CME with rates of the form \eqref{gen_noneq_rates} and $g_\rho=g$ for all reactions $\rho$, reads

\begin{align} \nn
 \sum_\rho \tilde{k}_\rho  g(\bm{n}-\bm{s}^\rho) e^{\beta[ F(\bm{n}-\bm{v}^{\rho})-F(\bm{n}-\bm{s}^\rho)+\sum_b r^\rho_b \mu_b]} P(\bm{n}-\bm{v}^{\rho}) +&\\\nn
 \sum_\rho \tilde{k}_\rho g(\bm{n}-\bm{r}^\rho) e^{\beta [F(\bm{n}+\bm{v}^\rho)-F(\bm{n}-\bm{r}^\rho)+\sum_b s^\rho_b \mu_b]}  P(\bm{n}+\bm{v}^{\rho})&=\\ 
 \sum_\rho \Big\{ \tilde{k}_\rho g(\bm{n}-\bm{r}^\rho) e^{\beta[ F(\bm{n})-F(\bm{n}-\bm{r}^{\rho})+\sum_b r^\rho_b \mu_b]} +&\\ \nn
\tilde{k}_\rho g(\bm{n}-\bm{s}^\rho) e^{\beta [F(\bm{n})-F(\bm{n}-\bm{s}^\rho )+\sum_b s^\rho_b \mu_b]}\Big\} P(\bm{n}) &.
\end{align}

By dividing the previous expression by $P(\bm{n})$ and substituting the ansatz (\ref{claim}) into it, we obtain
\begin{align}
\sum_\rho  \tilde{k}_\rho \Big \lbrace g(\bm{n}-\bm{s}^\rho) e^{\beta [ F(\bm{n}) -F(\bm{n}-\bm{s}^\rho)+\sum_a v_a^\rho \tilde{\mu}_a +\sum_b r^\rho_b \mu_b ] } +&\\ \nn
 g(\bm{n}-\bm{r}^\rho) e^{\beta [ F(\bm{n}) -F(\bm{n}-\bm{r}^\rho) -\sum_a v_a^\rho \tilde{\mu}_a +\sum_b s^\rho_b \mu_b] } \Big \rbrace &= \\\nn
 \sum_\rho \tilde{k}_\rho \Big \lbrace  g(\bm{n}-\bm{r}^\rho) e^{\beta [ F(\bm{n}) -F(\bm{n}-\bm{r}^\rho) +\sum_b r^\rho_b \mu_b] } +&\\\nn
  g(\bm{n}-\bm{s}^\rho) e^{\beta [ F(\bm{n}) -F(\bm{n}-\bm{s}^\rho)+\sum_b s^\rho_b \mu_b]  } \Big \rbrace &.
\end{align}

We now rewrite the relation above in terms of a summation over each of the complexes $\bm{z} \in \mathcal{C}$  separately

\begin{align}\nn
\sum_{\bm{z}} \Bigg \lbrace \sum_{\rho\vert\bm{s}^\rho=\bm{z} }  \tilde{k}_\rho  g(\bm{n}-\bm{s}^\rho)  e^{\beta [ F(\bm{n}) -F(\bm{n}-\bm{s}^\rho) +\sum_a v_a^\rho \tilde{\mu}_a +\sum_b r^\rho_b \mu_b] } +&\\ \nn \sum_{\rho\vert\bm{r}^\rho=\bm{z} } \tilde{k}_\rho g(\bm{n}-\bm{r}^\rho) e^{\beta [ F(\bm{n}) -F(\bm{n}-\bm{r}^\rho) -\sum_a v_a^\rho \tilde{\mu}_a +\sum_b s^\rho_b \mu_b ] } \Bigg \rbrace&= \\ 
 \sum_{\bm{z}} \Bigg \lbrace \sum_{\rho\vert\bm{r}^\rho=\bm{z}} \tilde{k}_\rho g(\bm{n}-\bm{r}^\rho) e^{\beta [ F(\bm{n}) -F(\bm{n}-\bm{r}^\rho) +\sum_b r^\rho_b \mu_b ] } +&\\ \nn
\sum_{\rho\vert\bm{s}^\rho=\bm{z} } \tilde{k}_\rho g(\bm{n}-\bm{s}^\rho) e^{\beta [ F(\bm{n}) -F(\bm{n}-\bm{s}^\rho)+\sum_b s^\rho_b \mu_b ]  } \Bigg \rbrace&,
\end{align}
where the subscript `$\rho\vert\bm{s}^\rho=\bm{z}$' denotes that the sum runs only over reactions $\rho$ whose product complex $\bm{s}^\rho$ is equal to $\bm{z}$. 
This previous equation will be satisfied if
\begin{align}\label{eq7}
\sum_{\rho\vert\bm{s}^\rho=\bm{z} }  \tilde{k}_\rho  g(\bm{n}-\bm{s}^\rho)  e^{\beta [ F(\bm{n}) -F(\bm{n}-\bm{s}^\rho) +\sum_a v_a^\rho \tilde{\mu}_a +\sum_b r^\rho_b \mu_b] } +&\\ \nn
 \sum_{\rho\vert\bm{r}^\rho=\bm{z} } \tilde{k}_\rho g(\bm{n}-\bm{r}^\rho) e^{\beta [ F(\bm{n}) -F(\bm{n}-\bm{r}^\rho) -\sum_a v_a^\rho \tilde{\mu}_a +\sum_b s^\rho_b \mu_b] } &= \\ \nn
 \sum_{\rho\vert\bm{r}^\rho=\bm{z} } \tilde{k}_\rho g(\bm{n}-\bm{r}^\rho) e^{\beta [ F(\bm{n}) -F(\bm{n}-\bm{r}^\rho)+\sum_b r^\rho_b \mu_b] } +\\ \nn
\sum_{\rho\vert\bm{s}^\rho=\bm{z} } \tilde{k}_\rho g(\bm{n}-\bm{s}^\rho) e^{\beta [ F(\bm{n}) -F(\bm{n}-\bm{s}^\rho)+\sum_b s^\rho_b \mu_b]  }&,
\end{align}
for every complex $\bm z$. For any given complex $\bm z$, \cref{eq7} can be rewritten in the following form:
\begin{align}
g(\bm{n}-\bm{z}) e^{\beta [ F(\bm{n}) -F(\bm{n}-\bm{z})]} \Bigg\lbrace \sum_{\rho\vert\bm{s}^\rho=\bm{z} }  \tilde{k}_\rho  e^{\beta [\sum_a v_a^\rho \tilde{\mu}_a +\sum_b r^\rho_b \mu_b]} +&\\ \nn
 \sum_{\rho\vert\bm{r}^\rho=\bm{z} } \tilde{k}_\rho  e^{\beta[ -\sum_a v_a^\rho \tilde{\mu}_a + \sum_b s^\rho_b \mu_b] } \Bigg\rbrace &= \\ \nn
 g(\bm{n}-\bm{z}) e^{\beta [ F(\bm{n}) -F(\bm{n}-\bm{z})]} \Bigg\lbrace \sum_{\rho\vert\bm{r}^\rho=\bm{z} } \tilde{k}_\rho e^{\beta \sum_b r^\rho_b \mu_b} +\sum_{\rho\vert\bm{s}^\rho=\bm{z} } \tilde{k}_\rho e^{\beta \sum_b s^\rho_b \mu_b } \Bigg\rbrace &.
\end{align}

We now divide both sides by $g(\bm{n}-\bm{z})  \exp \{\beta [ F(\bm{n}) -F(\bm{n}-\bm{z})]\}$, and obtain

\begin{align}\label{eq8}
\sum_{\rho\vert\bm{s}^\rho=\bm{z}}  \tilde{k}_\rho   e^{\beta [ \sum_a (z_a-r_a^\rho) \tilde{\mu}_a +\sum_b r^\rho_b \mu_b]} + \sum_{\rho\vert\bm{r}^\rho=\bm{z}} \tilde{k}_\rho e^{\beta [-\sum_a (s_a^\rho-z_a) \tilde{\mu}_a +\sum_b s^\rho_b \mu_b] }&=\\ \nn
\sum_{\rho\vert\bm{r}^\rho=\bm{z} } \tilde{k}_\rho e^{\beta \sum_b r^\rho_b \mu_b} +\sum_{\rho\vert\bm{s}^\rho=\bm{z} } \tilde{k}_\rho e^{\beta \sum_b s^\rho_b \mu_b }&,
\end{align}
where we have substituted $\bm{v}_\rho=\bm{s}_\rho-\bm{r}_\rho$ and, depending on the reactions over which the sum runs, one of this terms can be replaced by the complex $\bm{z}$. 

Finally, given that in \cref{eq8} $\bm{z}$ is fixed, we can divide both sides by $\exp (\beta \sum_a z_a \tilde{\mu}_a)$, yielding
\begin{align}\label{eq10}
\sum_{\rho\vert\bm{s}^\rho=\bm{z}}  \tilde{k}_\rho   e^{\beta [ -\sum_a r_a^\rho \tilde{\mu}_a +\sum_b r^\rho_b \mu_b]} + \sum_{\rho\vert\bm{r}^\rho=\bm{z}} \tilde{k}_\rho e^{\beta [-\sum_a s_a^\rho \tilde{\mu}_a +\sum_b s^\rho_b \mu_b] }&=\\ \nn
\sum_{\rho\vert\bm{r}^\rho=\bm{z} } \tilde{k}_\rho e^{\beta [\sum_b r^\rho_b \mu_b-\sum_a z_a^\rho \tilde{\mu}_a]} +\sum_{\rho\vert\bm{s}^\rho=\bm{z} } \tilde{k}_\rho e^{\beta [\sum_b s^\rho_b \mu_b-\sum_a z_a^\rho \tilde{\mu}_a] }&.
\end{align}

By inserting \cref{eq9a,k+,k-} in \cref{eq10},  we recover the complex-balance condition \eqref{CB_MAK} for an ideal and deterministic network. 

\section{Proof of the minimisation of the Lyapunov function} \label{proof_minimization}

In this Section we provide the full proof of \cref{der_L}. 

Since $Z$ does not depend on time, the time derivative of the Lyapunov function can be written as
\begin{align}\nn
\frac{d  \mathcal{L}}{dt}&=\\\nn
\sum_{k} \frac{\partial \mathcal{L}}{\partial c_{k}} \frac{\partial c_{k}}{\partial t} \nonumber &= \\
\beta \sum_{k} (\mu_{k}+\tilde{\mu}_{k}) \Big\{  \sum_\rho v_{k}^\rho k_\rho g(\bm{c}) \Big[ e^{\beta[ \sum_a r_a^\rho \mu_a+ \sum_b r_b^\rho \mu_b]} -e^{\beta [\sum_a s_a^\rho \mu_a+\sum_b s_b^\rho \mu_b]}\Big] \Big\}&,
\end{align}
where in the third line we have used \cref{deterministic_kinetics} with currents  given by \cref{gen_noneq_rates_det}. By adding and subtracting terms of the form $\sum_a r_a^\rho \tilde{\mu}_a$ in the exponentials (of the form $\sum_a s_a^\rho \tilde{\mu}_a$ for the second exponential), we  rewrite the previous equality as

\begin{align}\nn
\frac{d  \mathcal{L}}{dt}=&  \sum_\rho \sum_{k}  k_\rho g(\bm{c})  (\mu_{k}+\tilde{\mu}_{k}) (s_{k}^\rho-r_{k}^\rho) e^{\beta [ \sum_a r_a^\rho (\mu_a+\tilde{\mu}_a) - \sum_a r_a^\rho \tilde{\mu}_a + \sum_b r_b^\rho \mu_b]} \, + \\\label{eq11}
 &\sum_\rho \sum_{k}  k_\rho g(\bm{c})  (\mu_{k}+\tilde{\mu}_{k}) (r_{k}^\rho-s_{k}^\rho)  e^{\beta [\sum_a s_a^\rho (\mu_a+\tilde{\mu}_a) - \sum_a s_a^\rho \tilde{\mu}_a + \sum_b r_b^\rho \mu_b]}. 
\end{align}

We now consider   the inequality $e^s(t-s)\le e^t-e^s$---which results from $1+x \le e^x,\, \forall \,x\, \in \mathbb{R}$ with $x=t-s$---and apply it  to the sums of chemical potentials. In the first term in the RHS of \cref{eq11}, we set $s=\sum_a(\mu_a+\tilde{\mu}_a) r_a^\rho$ and $t=\sum_a(\mu_a+\tilde{\mu}_a) s_a^\rho$, and conversely in the second term. We then obtain

\begin{align}\nn
\frac{d  \mathcal{L}}{dt} \le & \sum_\rho  k_\rho g(\bm{c}) e^{\beta (\sum_b r_b^\rho \mu_b- \sum_a r_a^\rho \tilde{\mu}_a)} \left[ e^{ \beta \sum_a (\mu_a+\tilde{\mu}_a)s_a^\rho}- e^{\beta \sum_a r_a^\rho (\mu_a+\tilde{\mu}_a)}\right]+ \\ 
 & \sum_\rho k_\rho g(\bm{c}) e^{\beta (\sum_b s_b^\rho \mu_b- \sum_a s_a^\rho \tilde{\mu}_a) }  \left[  e^{\beta  \sum_a(\mu_a+\tilde{\mu}_a)r_a^\rho}- e^{\beta \sum_a s_a^\rho (\mu_a+\tilde{\mu}_a)}\right].
\end{align}

This expression can now be separated in terms of the different complexes in the system:
\begin{align} \label{eqC18}
\frac{d  \mathcal{L}}{dt} \le &\sum_{\bm{z} \in \mathcal{C}} g(\bm{c}) \Bigg \lbrace \sum_{\rho\vert\bm{s}^\rho=\bm{z}} k_\rho  e^{\beta [\sum_b r_b^\rho \mu_b-\sum_a r_a^\rho \tilde{\mu}_a+ \sum_a (\mu_a+\tilde{\mu}_a)s_a^\rho]}-\\\nn
&\sum_{\rho\vert\bm{r}^\rho=\bm{z}} k_\rho  e^{\beta [\sum_b r_b^\rho \mu_b-\sum_a r_a^\rho \tilde{\mu}_a+\beta \sum_a r_a^\rho (\mu_a+\tilde{\mu}_a)]}+ \\\nonumber
 & \sum_{\rho\vert\bm{r}^\rho=\bm{z}}  k_\rho e^{\beta [\sum_b s_b^\rho \mu_b-\sum_a s_a^\rho \tilde{\mu}_a +   \sum_a(\mu_a+\tilde{\mu}_a)r_a^\rho]}-\\ \nn
& \sum_{\rho\vert\bm{s}^\rho=\bm{z}}  k_\rho e^{\beta [\sum_b s_b^\rho \mu_b- \sum_a s_a^\rho \tilde{\mu}_a+ \sum_a s_a^\rho (\mu_a+\tilde{\mu}_a)]}\Bigg \rbrace.
\end{align}

Now, for a complex-balance system with MAK, we know that
\begin{align}
\label{C19}
\sum_{\rho\vert\bm{s}^\rho=\bm{z}} k_\rho  e^{\beta (\sum_b r_b^\rho \mu_b-\sum_a r_a^\rho \tilde{\mu}_a)}-\sum_{\rho\vert\bm{r}^\rho=\bm{z}} k_\rho  e^{\beta (\sum_b r_b^\rho \mu_b-\sum_a r_a^\rho \tilde{\mu}_a)}\, + &\\ \nn
 \sum_{\rho\vert\bm{r}^\rho=\bm{z}}  k_\rho e^{\beta (\sum_b s_b^\rho \mu_b-\sum_a s_a^\rho \tilde{\mu}_a)}- \sum_{\rho\vert\bm{s}^\rho=\bm{z}}  k_\rho e^{\beta (\sum_b s_b^\rho \mu_b-\sum_a s_a^\rho \tilde{\mu}_a)}&=0,
\end{align}
for all complex $\mathbf{z}$, see \cref{eq10}. Multiplying \cref{C19} by $\exp[\beta\sum_a z_a( \mu_a+\tilde{\mu}_a)]$ (since the complex $\mathbf{z}$ is fixed), we obtain an equality for each  complex $\bm{z}$:
\begin{align}
& \sum_{\rho\vert\bm{s}^\rho=\bm{z}} k_\rho  e^{\beta [\sum_b r_b^\rho \mu_b-\sum_a r_a^\rho \tilde{\mu}_a+ \sum_a (\mu_a+\tilde{\mu}_a)s_a^\rho]}-\\\nn
&\sum_{\rho\vert\bm{r}^\rho=\bm{z}} k_\rho  e^{\beta [\sum_b r_b^\rho \mu_b-\sum_a r_a^\rho \tilde{\mu}_a+\beta \sum_a r_a^\rho (\mu_a+\tilde{\mu}_a)]}+ \\\nonumber
 & \sum_{\rho\vert\bm{r}^\rho=\bm{z}}  k_\rho e^{\beta [\sum_b s_b^\rho \mu_b-\sum_a s_a^\rho \tilde{\mu}_a +   \sum_a(\mu_a+\tilde{\mu}_a)r_a^\rho]}-\\ \nn
& \sum_{\rho\vert\bm{s}^\rho=\bm{z}}  k_\rho e^{\beta [\sum_b s_b^\rho \mu_b- \sum_a s_a^\rho \tilde{\mu}_a+ \sum_a s_a^\rho (\mu_a+\tilde{\mu}_a)]}=0,
\end{align}
which is precisely the term in brackets in \cref{eqC18}. Summing over all complexes yields the inequality \eqref{der_L}.


\section{Minimisation of the Lagrangian to obtain the phase diagram} \label{minimization}

In order to find the steady state of the system, we need to minimise the Lyapunov functional or,  in the presence of particle-conservation constraints, the Lagrangian, \eqref{Lyap_functional}. 
A substantial simplification can be made by neglecting the contribution of the interfaces, i.e.,  considering the system as composed of two homogeneous phases. In this approximation, the function which needs to be minimised is the Lagrangian  \eqref{Lagrangian2}, which depends on eight independent variables: $\phi_a^p$ for $a=\rm A,B,C$, $ p=1,2$, $\lambda$ and $\vert\Omega_1\vert$. 

First, we reduce the dimensionality of the problem by equating the derivatives of the Lagrangian with respect to the concentrations of the species:
\begin{equation}\label{eq12}
\frac{\partial \Lambda}{\partial \phi_a^p}=\frac{\partial \Lambda}{\partial  \phi_{k}^p},
\end{equation}
where $a$ and $k$ denote two chemical species, and $p = 1,2$ refers to the phases. 
\Cref{eq12} for a system at equilibrium yields the equality of chemical potentials (with their appropriate stoichiometry). Here, however, \Cref{eq12} includes the shifted chemical potential term $\tilde{\mu}$ due to the out-of-equilibrium complex-balancing. For a simple free energy like \cref{free_phiA}, \cref{eq12} implies
\begin{align}
\phi_{\rm B}^p=&\phi_{\rm A}^p e^{2\chi\phi_{\rm A}^p+\tilde{\mu}_{\rm A}- \tilde{\mu}_{\rm B}} , \\
\phi_{\rm C}^p=&\phi_{\rm A}^p e^{2\chi\phi_{\rm A}^p+\tilde{\mu}_{\rm A}- \tilde{\mu}_{\rm C}} ,
\end{align}
which reduces the problem to just four variables: $\phi^1_{\rm A}, \, \phi^2_{\rm A}$, $\lambda$ and $\vert\Omega^1\vert$.
     
Finally, given that we are interested in the phase diagram of the mixture and not in the actual steady state of the solution (i.e., we do not need to know how much volume each of the phases occupies), we can avoid solving for $\vert\Omega^1\vert$.  This can be achieved by enforcing the stationarity condition of the Lagrangian with respect to the volume:
\begin{equation}
\frac{\partial \Lambda}{\partial \vert\Omega^1\vert}=0,
\end{equation}
which, together with
\begin{align}
\frac{\partial \Lambda}{\partial \phi^1_{\rm A}}=0 , \\
\frac{\partial \Lambda}{\partial \phi^2_{\rm A}}=0 ,
\end{align}
yields a fully determined system for the unknowns $\phi^1_{\rm A}, \, \phi^2_{\rm A}$ and $\lambda$ (the dependency on $\vert\Omega^1\vert$ drops out). The resulting equations for such unknowns are transcendental equations which, in general, have no explicit analytical solution. Therefore, they need to be solved numerically. Even the numerical solution is involved as the parameters near criticality, which is why in Fig. \ref{phase_diagram} the density of data around the critical point decreases.

\end{appendices}

\bibliography{bibliography_stocCRN_phase_sep.bib}

\begin{thebibliography}{51}%
\makeatletter
\providecommand \@ifxundefined [1]{%
 \@ifx{#1\undefined}
}%
\providecommand \@ifnum [1]{%
 \ifnum #1\expandafter \@firstoftwo
 \else \expandafter \@secondoftwo
 \fi
}%
\providecommand \@ifx [1]{%
 \ifx #1\expandafter \@firstoftwo
 \else \expandafter \@secondoftwo
 \fi
}%
\providecommand \natexlab [1]{#1}%
\providecommand \enquote  [1]{``#1''}%
\providecommand \bibnamefont  [1]{#1}%
\providecommand \bibfnamefont [1]{#1}%
\providecommand \citenamefont [1]{#1}%
\providecommand \href@noop [0]{\@secondoftwo}%
\providecommand \href [0]{\begingroup \@sanitize@url \@href}%
\providecommand \@href[1]{\@@startlink{#1}\@@href}%
\providecommand \@@href[1]{\endgroup#1\@@endlink}%
\providecommand \@sanitize@url [0]{\catcode `\\12\catcode `\$12\catcode
  `\&12\catcode `\#12\catcode `\^12\catcode `\_12\catcode `\%12\relax}%
\providecommand \@@startlink[1]{}%
\providecommand \@@endlink[0]{}%
\providecommand \url  [0]{\begingroup\@sanitize@url \@url }%
\providecommand \@url [1]{\endgroup\@href {#1}{\urlprefix }}%
\providecommand \urlprefix  [0]{URL }%
\providecommand \Eprint [0]{\href }%
\providecommand \doibase [0]{https://doi.org/}%
\providecommand \selectlanguage [0]{\@gobble}%
\providecommand \bibinfo  [0]{\@secondoftwo}%
\providecommand \bibfield  [0]{\@secondoftwo}%
\providecommand \translation [1]{[#1]}%
\providecommand \BibitemOpen [0]{}%
\providecommand \bibitemStop [0]{}%
\providecommand \bibitemNoStop [0]{.\EOS\space}%
\providecommand \EOS [0]{\spacefactor3000\relax}%
\providecommand \BibitemShut  [1]{\csname bibitem#1\endcsname}%
\let\auto@bib@innerbib\@empty
\bibitem [{\citenamefont {Zielinski}\ \emph {et~al.}(2017)\citenamefont
  {Zielinski}, \citenamefont {McGann}, \citenamefont {Nychka},\ and\
  \citenamefont {Elliott}}]{Zielinski2017Nonideal}%
  \BibitemOpen
  \bibfield  {author} {\bibinfo {author} {\bibfnamefont {M.~W.}\ \bibnamefont
  {Zielinski}}, \bibinfo {author} {\bibfnamefont {L.~E.}\ \bibnamefont
  {McGann}}, \bibinfo {author} {\bibfnamefont {J.~A.}\ \bibnamefont {Nychka}},\
  and\ \bibinfo {author} {\bibfnamefont {J.~A.~W.}\ \bibnamefont {Elliott}},\
  }\bibfield  {title} {\bibinfo {title} {Nonideal solute chemical potential
  equation and the validity of the grouped solute approach for intracellular
  solution thermodynamics},\ }\href@noop {} {\bibfield  {journal} {\bibinfo
  {journal} {The Journal of Physical Chemistry B}\ }\textbf {\bibinfo {volume}
  {121}},\ \bibinfo {pages} {10443} (\bibinfo {year} {2017})}\BibitemShut
  {NoStop}%
\bibitem [{\citenamefont {Mitchison}(2019)}]{Mitchison2019ColloidCrowding}%
  \BibitemOpen
  \bibfield  {author} {\bibinfo {author} {\bibfnamefont {T.~J.}\ \bibnamefont
  {Mitchison}},\ }\bibfield  {title} {\bibinfo {title} {Colloid osmotic
  parameterization and measurement of subcellular crowding},\ }\href@noop {}
  {\bibfield  {journal} {\bibinfo  {journal} {Molecular Biology of the Cell}\
  }\textbf {\bibinfo {volume} {30}},\ \bibinfo {pages} {173} (\bibinfo {year}
  {2019})}\BibitemShut {NoStop}%
\bibitem [{\citenamefont {Zhou}\ \emph {et~al.}(2008)\citenamefont {Zhou},
  \citenamefont {Rivas},\ and\ \citenamefont {Minton}}]{Minton2008Crowding}%
  \BibitemOpen
  \bibfield  {author} {\bibinfo {author} {\bibfnamefont {H.-X.}\ \bibnamefont
  {Zhou}}, \bibinfo {author} {\bibfnamefont {G.}~\bibnamefont {Rivas}},\ and\
  \bibinfo {author} {\bibfnamefont {A.~P.}\ \bibnamefont {Minton}},\ }\bibfield
   {title} {\bibinfo {title} {Macromolecular crowding and confinement:
  Biochemical, biophysical, and potential physiological consequences},\
  }\href@noop {} {\bibfield  {journal} {\bibinfo  {journal} {Annual Review of
  Biophysics}\ }\textbf {\bibinfo {volume} {37}},\ \bibinfo {pages} {375}
  (\bibinfo {year} {2008})}\BibitemShut {NoStop}%
\bibitem [{\citenamefont {Fall}\ and\ \citenamefont
  {Keizer}(2001)}]{Fall2001MitochondrialCa2+}%
  \BibitemOpen
  \bibfield  {author} {\bibinfo {author} {\bibfnamefont {C.~P.}\ \bibnamefont
  {Fall}}\ and\ \bibinfo {author} {\bibfnamefont {J.~E.}\ \bibnamefont
  {Keizer}},\ }\bibfield  {title} {\bibinfo {title} {Mitochondrial modulation
  of intracellular {Ca2}+ signaling},\ }\href@noop {} {\bibfield  {journal}
  {\bibinfo  {journal} {Journal of Theoretical Biology}\ }\textbf {\bibinfo
  {volume} {210}},\ \bibinfo {pages} {151} (\bibinfo {year}
  {2001})}\BibitemShut {NoStop}%
\bibitem [{\citenamefont {Wang}\ \emph {et~al.}(2018)\citenamefont {Wang},
  \citenamefont {Choi}, \citenamefont {Holehouse}, \citenamefont {Lee},
  \citenamefont {Zhang}, \citenamefont {Jahnel}, \citenamefont {Maharana},
  \citenamefont {Lemaitre}, \citenamefont {Pozniakovsky}, \citenamefont
  {Drechsel}, \citenamefont {Poser}, \citenamefont {Pappu}, \citenamefont
  {Alberti},\ and\ \citenamefont {Hyman}}]{Hyman2018Grammar}%
  \BibitemOpen
  \bibfield  {author} {\bibinfo {author} {\bibfnamefont {J.}~\bibnamefont
  {Wang}}, \bibinfo {author} {\bibfnamefont {J.-M.}\ \bibnamefont {Choi}},
  \bibinfo {author} {\bibfnamefont {A.~S.}\ \bibnamefont {Holehouse}}, \bibinfo
  {author} {\bibfnamefont {H.~O.}\ \bibnamefont {Lee}}, \bibinfo {author}
  {\bibfnamefont {X.}~\bibnamefont {Zhang}}, \bibinfo {author} {\bibfnamefont
  {M.}~\bibnamefont {Jahnel}}, \bibinfo {author} {\bibfnamefont
  {S.}~\bibnamefont {Maharana}}, \bibinfo {author} {\bibfnamefont
  {R.}~\bibnamefont {Lemaitre}}, \bibinfo {author} {\bibfnamefont
  {A.}~\bibnamefont {Pozniakovsky}}, \bibinfo {author} {\bibfnamefont
  {D.}~\bibnamefont {Drechsel}}, \bibinfo {author} {\bibfnamefont
  {I.}~\bibnamefont {Poser}}, \bibinfo {author} {\bibfnamefont {R.~V.}\
  \bibnamefont {Pappu}}, \bibinfo {author} {\bibfnamefont {S.}~\bibnamefont
  {Alberti}},\ and\ \bibinfo {author} {\bibfnamefont {A.~A.}\ \bibnamefont
  {Hyman}},\ }\bibfield  {title} {\bibinfo {title} {A molecular grammar
  governing the driving forces for phase separation of prion-like {RNA} binding
  proteins},\ }\href@noop {} {\bibfield  {journal} {\bibinfo  {journal} {Cell}\
  }\textbf {\bibinfo {volume} {174}},\ \bibinfo {pages} {688} (\bibinfo {year}
  {2018})}\BibitemShut {NoStop}%
\bibitem [{\citenamefont {Brangwynne}\ \emph {et~al.}(2009)\citenamefont
  {Brangwynne}, \citenamefont {Eckmann}, \citenamefont {Courson}, \citenamefont
  {Rybarska}, \citenamefont {Hoege}, \citenamefont {Gharakhani}, \citenamefont
  {J{\"u}licher},\ and\ \citenamefont {Hyman}}]{brangwynne2009germline}%
  \BibitemOpen
  \bibfield  {author} {\bibinfo {author} {\bibfnamefont {C.~P.}\ \bibnamefont
  {Brangwynne}}, \bibinfo {author} {\bibfnamefont {C.~R.}\ \bibnamefont
  {Eckmann}}, \bibinfo {author} {\bibfnamefont {D.~S.}\ \bibnamefont
  {Courson}}, \bibinfo {author} {\bibfnamefont {A.}~\bibnamefont {Rybarska}},
  \bibinfo {author} {\bibfnamefont {C.}~\bibnamefont {Hoege}}, \bibinfo
  {author} {\bibfnamefont {J.}~\bibnamefont {Gharakhani}}, \bibinfo {author}
  {\bibfnamefont {F.}~\bibnamefont {J{\"u}licher}},\ and\ \bibinfo {author}
  {\bibfnamefont {A.~A.}\ \bibnamefont {Hyman}},\ }\bibfield  {title} {\bibinfo
  {title} {Germline {P} granules are liquid droplets that localize by
  controlled dissolution/condensation},\ }\href@noop {} {\bibfield  {journal}
  {\bibinfo  {journal} {Science}\ }\textbf {\bibinfo {volume} {324}},\ \bibinfo
  {pages} {1729} (\bibinfo {year} {2009})}\BibitemShut {NoStop}%
\bibitem [{\citenamefont {Li}\ \emph {et~al.}(2012)\citenamefont {Li},
  \citenamefont {Banjade}, \citenamefont {Cheng}, \citenamefont {Kim},
  \citenamefont {Chen}, \citenamefont {Guo}, \citenamefont {Llaguno},
  \citenamefont {Hollingsworth}, \citenamefont {King}, \citenamefont {Banani},
  \citenamefont {Russo}, \citenamefont {Jiang}, \citenamefont {Nixon},\ and\
  \citenamefont {Rosen}}]{Rosen2012PhaseTrans}%
  \BibitemOpen
  \bibfield  {author} {\bibinfo {author} {\bibfnamefont {P.}~\bibnamefont
  {Li}}, \bibinfo {author} {\bibfnamefont {S.}~\bibnamefont {Banjade}},
  \bibinfo {author} {\bibfnamefont {H.-C.}\ \bibnamefont {Cheng}}, \bibinfo
  {author} {\bibfnamefont {S.}~\bibnamefont {Kim}}, \bibinfo {author}
  {\bibfnamefont {B.}~\bibnamefont {Chen}}, \bibinfo {author} {\bibfnamefont
  {L.}~\bibnamefont {Guo}}, \bibinfo {author} {\bibfnamefont {M.}~\bibnamefont
  {Llaguno}}, \bibinfo {author} {\bibfnamefont {J.~V.}\ \bibnamefont
  {Hollingsworth}}, \bibinfo {author} {\bibfnamefont {D.~S.}\ \bibnamefont
  {King}}, \bibinfo {author} {\bibfnamefont {S.~F.}\ \bibnamefont {Banani}},
  \bibinfo {author} {\bibfnamefont {P.~S.}\ \bibnamefont {Russo}}, \bibinfo
  {author} {\bibfnamefont {Q.-X.}\ \bibnamefont {Jiang}}, \bibinfo {author}
  {\bibfnamefont {B.~T.}\ \bibnamefont {Nixon}},\ and\ \bibinfo {author}
  {\bibfnamefont {M.~K.}\ \bibnamefont {Rosen}},\ }\bibfield  {title} {\bibinfo
  {title} {Phase transitions in the assembly of multivalent signalling
  proteins},\ }\href@noop {} {\bibfield  {journal} {\bibinfo  {journal}
  {Nature}\ }\textbf {\bibinfo {volume} {483}},\ \bibinfo {pages} {336}
  (\bibinfo {year} {2012})}\BibitemShut {NoStop}%
\bibitem [{\citenamefont {Su}\ \emph {et~al.}(2016)\citenamefont {Su},
  \citenamefont {Ditlev}, \citenamefont {Hui}, \citenamefont {Xing},
  \citenamefont {Banjade}, \citenamefont {Okrut}, \citenamefont {King},
  \citenamefont {Taunton}, \citenamefont {Rosen},\ and\ \citenamefont
  {Vale}}]{Vale2016PhaseSep}%
  \BibitemOpen
  \bibfield  {author} {\bibinfo {author} {\bibfnamefont {X.}~\bibnamefont
  {Su}}, \bibinfo {author} {\bibfnamefont {J.~A.}\ \bibnamefont {Ditlev}},
  \bibinfo {author} {\bibfnamefont {E.}~\bibnamefont {Hui}}, \bibinfo {author}
  {\bibfnamefont {W.}~\bibnamefont {Xing}}, \bibinfo {author} {\bibfnamefont
  {S.}~\bibnamefont {Banjade}}, \bibinfo {author} {\bibfnamefont
  {J.}~\bibnamefont {Okrut}}, \bibinfo {author} {\bibfnamefont {D.~S.}\
  \bibnamefont {King}}, \bibinfo {author} {\bibfnamefont {J.}~\bibnamefont
  {Taunton}}, \bibinfo {author} {\bibfnamefont {M.~K.}\ \bibnamefont {Rosen}},\
  and\ \bibinfo {author} {\bibfnamefont {R.~D.}\ \bibnamefont {Vale}},\
  }\bibfield  {title} {\bibinfo {title} {Phase separation of signaling
  molecules promotes {T} cell receptor signal transduction},\ }\href@noop {}
  {\bibfield  {journal} {\bibinfo  {journal} {Science}\ }\textbf {\bibinfo
  {volume} {352}},\ \bibinfo {pages} {595} (\bibinfo {year}
  {2016})}\BibitemShut {NoStop}%
\bibitem [{\citenamefont {Ditlev}\ \emph {et~al.}(2018)\citenamefont {Ditlev},
  \citenamefont {Case},\ and\ \citenamefont {Rosen}}]{Rosen2018Whosin}%
  \BibitemOpen
  \bibfield  {author} {\bibinfo {author} {\bibfnamefont {J.~A.}\ \bibnamefont
  {Ditlev}}, \bibinfo {author} {\bibfnamefont {L.~B.}\ \bibnamefont {Case}},\
  and\ \bibinfo {author} {\bibfnamefont {M.~K.}\ \bibnamefont {Rosen}},\
  }\bibfield  {title} {\bibinfo {title} {Who's in and who's out—compositional
  control of biomolecular condensates},\ }\href@noop {} {\bibfield  {journal}
  {\bibinfo  {journal} {Journal of Molecular Biology}\ }\textbf {\bibinfo
  {volume} {430}},\ \bibinfo {pages} {4666} (\bibinfo {year}
  {2018})}\BibitemShut {NoStop}%
\bibitem [{\citenamefont {Banani}\ \emph {et~al.}(2017)\citenamefont {Banani},
  \citenamefont {Lee}, \citenamefont {Hyman},\ and\ \citenamefont
  {Rosen}}]{Banani2017Review}%
  \BibitemOpen
  \bibfield  {author} {\bibinfo {author} {\bibfnamefont {S.~F.}\ \bibnamefont
  {Banani}}, \bibinfo {author} {\bibfnamefont {H.~O.}\ \bibnamefont {Lee}},
  \bibinfo {author} {\bibfnamefont {A.~A.}\ \bibnamefont {Hyman}},\ and\
  \bibinfo {author} {\bibfnamefont {M.~K.}\ \bibnamefont {Rosen}},\ }\bibfield
  {title} {\bibinfo {title} {Biomolecular condensates: organizers of cellular
  biochemistry},\ }\href@noop {} {\bibfield  {journal} {\bibinfo  {journal}
  {Nature Reviews Molecular Cell Biology}\ }\textbf {\bibinfo {volume} {18}},\
  \bibinfo {pages} {285} (\bibinfo {year} {2017})}\BibitemShut {NoStop}%
\bibitem [{\citenamefont {Castellana}\ \emph {et~al.}(2014)\citenamefont
  {Castellana}, \citenamefont {Wilson}, \citenamefont {Xu}, \citenamefont
  {Joshi}, \citenamefont {Cristea}, \citenamefont {Rabinowitz}, \citenamefont
  {Gitai},\ and\ \citenamefont {Wingreen}}]{castellana2014enzyme}%
  \BibitemOpen
  \bibfield  {author} {\bibinfo {author} {\bibfnamefont {M.}~\bibnamefont
  {Castellana}}, \bibinfo {author} {\bibfnamefont {M.~Z.}\ \bibnamefont
  {Wilson}}, \bibinfo {author} {\bibfnamefont {Y.}~\bibnamefont {Xu}}, \bibinfo
  {author} {\bibfnamefont {P.}~\bibnamefont {Joshi}}, \bibinfo {author}
  {\bibfnamefont {I.~M.}\ \bibnamefont {Cristea}}, \bibinfo {author}
  {\bibfnamefont {J.~D.}\ \bibnamefont {Rabinowitz}}, \bibinfo {author}
  {\bibfnamefont {Z.}~\bibnamefont {Gitai}},\ and\ \bibinfo {author}
  {\bibfnamefont {N.~S.}\ \bibnamefont {Wingreen}},\ }\bibfield  {title}
  {\bibinfo {title} {Enzyme clustering accelerates processing of intermediates
  through metabolic channeling},\ }\href@noop {} {\bibfield  {journal}
  {\bibinfo  {journal} {Nature {B}iotechnology}\ }\textbf {\bibinfo {volume}
  {32}},\ \bibinfo {pages} {1011} (\bibinfo {year} {2014})}\BibitemShut
  {NoStop}%
\bibitem [{\citenamefont {Vagne}\ \emph {et~al.}(2015)\citenamefont {Vagne},
  \citenamefont {Turner},\ and\ \citenamefont {Sens}}]{vagne2015sensing}%
  \BibitemOpen
  \bibfield  {author} {\bibinfo {author} {\bibfnamefont {Q.}~\bibnamefont
  {Vagne}}, \bibinfo {author} {\bibfnamefont {M.~S.}\ \bibnamefont {Turner}},\
  and\ \bibinfo {author} {\bibfnamefont {P.}~\bibnamefont {Sens}},\ }\bibfield
  {title} {\bibinfo {title} {Sensing size through clustering in non-equilibrium
  membranes and the control of membrane-bound enzymatic reactions},\
  }\href@noop {} {\bibfield  {journal} {\bibinfo  {journal} {PloS one}\
  }\textbf {\bibinfo {volume} {10}},\ \bibinfo {pages} {e0143470} (\bibinfo
  {year} {2015})}\BibitemShut {NoStop}%
\bibitem [{\citenamefont {Buchner}\ \emph {et~al.}(2013)\citenamefont
  {Buchner}, \citenamefont {Tostevin}, \citenamefont {Hinzpeter},\ and\
  \citenamefont {Gerland}}]{buchner2013optimization}%
  \BibitemOpen
  \bibfield  {author} {\bibinfo {author} {\bibfnamefont {A.}~\bibnamefont
  {Buchner}}, \bibinfo {author} {\bibfnamefont {F.}~\bibnamefont {Tostevin}},
  \bibinfo {author} {\bibfnamefont {F.}~\bibnamefont {Hinzpeter}},\ and\
  \bibinfo {author} {\bibfnamefont {U.}~\bibnamefont {Gerland}},\ }\bibfield
  {title} {\bibinfo {title} {Optimization of collective enzyme activity via
  spatial localization},\ }\href@noop {} {\bibfield  {journal} {\bibinfo
  {journal} {Journal of Chemical Physics}\ }\textbf {\bibinfo {volume} {139}},\
  \bibinfo {pages} {135101} (\bibinfo {year} {2013})}\BibitemShut {NoStop}%
\bibitem [{\citenamefont {Tsang}\ \emph {et~al.}(2019)\citenamefont {Tsang},
  \citenamefont {Arsenault}, \citenamefont {Vernon}, \citenamefont {Lin},
  \citenamefont {Sonenberg}, \citenamefont {Wang}, \citenamefont {Bah},\ and\
  \citenamefont {Forman-Kay}}]{FormanKay2019Phosphoregulated}%
  \BibitemOpen
  \bibfield  {author} {\bibinfo {author} {\bibfnamefont {B.}~\bibnamefont
  {Tsang}}, \bibinfo {author} {\bibfnamefont {J.}~\bibnamefont {Arsenault}},
  \bibinfo {author} {\bibfnamefont {R.~M.}\ \bibnamefont {Vernon}}, \bibinfo
  {author} {\bibfnamefont {H.}~\bibnamefont {Lin}}, \bibinfo {author}
  {\bibfnamefont {N.}~\bibnamefont {Sonenberg}}, \bibinfo {author}
  {\bibfnamefont {L.-Y.}\ \bibnamefont {Wang}}, \bibinfo {author}
  {\bibfnamefont {A.}~\bibnamefont {Bah}},\ and\ \bibinfo {author}
  {\bibfnamefont {J.~D.}\ \bibnamefont {Forman-Kay}},\ }\bibfield  {title}
  {\bibinfo {title} {Phosphoregulated {FMRP} phase separation models
  activity-dependent translation through bidirectional control of mrna granule
  formation},\ }\href@noop {} {\bibfield  {journal} {\bibinfo  {journal}
  {Proceedings of the National Academy of Sciences}\ }\textbf {\bibinfo
  {volume} {116}},\ \bibinfo {pages} {4218} (\bibinfo {year}
  {2019})}\BibitemShut {NoStop}%
\bibitem [{\citenamefont {Kim}\ \emph {et~al.}(2019)\citenamefont {Kim},
  \citenamefont {Tsang}, \citenamefont {Vernon}, \citenamefont {Sonenberg},
  \citenamefont {Kay},\ and\ \citenamefont
  {Forman-Kay}}]{FormanKay2019Phosphodependent}%
  \BibitemOpen
  \bibfield  {author} {\bibinfo {author} {\bibfnamefont {T.~H.}\ \bibnamefont
  {Kim}}, \bibinfo {author} {\bibfnamefont {B.}~\bibnamefont {Tsang}}, \bibinfo
  {author} {\bibfnamefont {R.~M.}\ \bibnamefont {Vernon}}, \bibinfo {author}
  {\bibfnamefont {N.}~\bibnamefont {Sonenberg}}, \bibinfo {author}
  {\bibfnamefont {L.~E.}\ \bibnamefont {Kay}},\ and\ \bibinfo {author}
  {\bibfnamefont {J.~D.}\ \bibnamefont {Forman-Kay}},\ }\bibfield  {title}
  {\bibinfo {title} {Phospho-dependent phase separation of {FMRP} and {CAPRIN1}
  recapitulates regulation of translation and deadenylation},\ }\href@noop {}
  {\bibfield  {journal} {\bibinfo  {journal} {Science}\ }\textbf {\bibinfo
  {volume} {365}},\ \bibinfo {pages} {825} (\bibinfo {year}
  {2019})}\BibitemShut {NoStop}%
\bibitem [{\citenamefont {Nott}\ \emph {et~al.}(2015)\citenamefont {Nott},
  \citenamefont {Petsalaki}, \citenamefont {Farber}, \citenamefont {Jervis},
  \citenamefont {Fussner}, \citenamefont {Plochowietz}, \citenamefont {Craggs},
  \citenamefont {Bazett-Jones}, \citenamefont {Pawson}, \citenamefont
  {Forman-Kay},\ and\ \citenamefont {Baldwin}}]{FormanKay2016Nuage}%
  \BibitemOpen
  \bibfield  {author} {\bibinfo {author} {\bibfnamefont {T.}~\bibnamefont
  {Nott}}, \bibinfo {author} {\bibfnamefont {E.}~\bibnamefont {Petsalaki}},
  \bibinfo {author} {\bibfnamefont {P.}~\bibnamefont {Farber}}, \bibinfo
  {author} {\bibfnamefont {D.}~\bibnamefont {Jervis}}, \bibinfo {author}
  {\bibfnamefont {E.}~\bibnamefont {Fussner}}, \bibinfo {author} {\bibfnamefont
  {A.}~\bibnamefont {Plochowietz}}, \bibinfo {author} {\bibfnamefont {T.~D.}\
  \bibnamefont {Craggs}}, \bibinfo {author} {\bibfnamefont {D.}~\bibnamefont
  {Bazett-Jones}}, \bibinfo {author} {\bibfnamefont {T.}~\bibnamefont
  {Pawson}}, \bibinfo {author} {\bibfnamefont {J.}~\bibnamefont {Forman-Kay}},\
  and\ \bibinfo {author} {\bibfnamefont {A.}~\bibnamefont {Baldwin}},\
  }\bibfield  {title} {\bibinfo {title} {Phase transition of a disordered nuage
  protein generates environmentally responsive membraneless organelles},\
  }\href@noop {} {\bibfield  {journal} {\bibinfo  {journal} {Molecular Cell}\
  }\textbf {\bibinfo {volume} {57}},\ \bibinfo {pages} {936} (\bibinfo {year}
  {2015})}\BibitemShut {NoStop}%
\bibitem [{\citenamefont {Franzmann}\ \emph {et~al.}(2018)\citenamefont
  {Franzmann}, \citenamefont {Jahnel}, \citenamefont {Pozniakovsky},
  \citenamefont {Mahamid}, \citenamefont {Holehouse}, \citenamefont
  {N{\"u}ske}, \citenamefont {Richter}, \citenamefont {Baumeister},
  \citenamefont {Grill}, \citenamefont {Pappu}, \citenamefont {Hyman},\ and\
  \citenamefont {Alberti}}]{Hyman2018PhYeast}%
  \BibitemOpen
  \bibfield  {author} {\bibinfo {author} {\bibfnamefont {T.~M.}\ \bibnamefont
  {Franzmann}}, \bibinfo {author} {\bibfnamefont {M.}~\bibnamefont {Jahnel}},
  \bibinfo {author} {\bibfnamefont {A.}~\bibnamefont {Pozniakovsky}}, \bibinfo
  {author} {\bibfnamefont {J.}~\bibnamefont {Mahamid}}, \bibinfo {author}
  {\bibfnamefont {A.~S.}\ \bibnamefont {Holehouse}}, \bibinfo {author}
  {\bibfnamefont {E.}~\bibnamefont {N{\"u}ske}}, \bibinfo {author}
  {\bibfnamefont {D.}~\bibnamefont {Richter}}, \bibinfo {author} {\bibfnamefont
  {W.}~\bibnamefont {Baumeister}}, \bibinfo {author} {\bibfnamefont {S.~W.}\
  \bibnamefont {Grill}}, \bibinfo {author} {\bibfnamefont {R.~V.}\ \bibnamefont
  {Pappu}}, \bibinfo {author} {\bibfnamefont {A.~A.}\ \bibnamefont {Hyman}},\
  and\ \bibinfo {author} {\bibfnamefont {S.}~\bibnamefont {Alberti}},\
  }\bibfield  {title} {\bibinfo {title} {Phase separation of a yeast prion
  protein promotes cellular fitness},\ }\href@noop {} {\bibfield  {journal}
  {\bibinfo  {journal} {Science}\ }\textbf {\bibinfo {volume} {359}} (\bibinfo
  {year} {2018})}\BibitemShut {NoStop}%
\bibitem [{\citenamefont {Lyon}\ \emph {et~al.}(2021)\citenamefont {Lyon},
  \citenamefont {Peeples},\ and\ \citenamefont {Rosen}}]{Rosen2021Functions}%
  \BibitemOpen
  \bibfield  {author} {\bibinfo {author} {\bibfnamefont {A.~S.}\ \bibnamefont
  {Lyon}}, \bibinfo {author} {\bibfnamefont {W.~B.}\ \bibnamefont {Peeples}},\
  and\ \bibinfo {author} {\bibfnamefont {M.~K.}\ \bibnamefont {Rosen}},\
  }\bibfield  {title} {\bibinfo {title} {A framework for understanding the
  functions of biomolecular condensates across scales},\ }\href@noop {}
  {\bibfield  {journal} {\bibinfo  {journal} {Nature Reviews Molecular Cell
  Biology}\ }\textbf {\bibinfo {volume} {22}},\ \bibinfo {pages} {215}
  (\bibinfo {year} {2021})}\BibitemShut {NoStop}%
\bibitem [{\citenamefont {Huberman}(1976)}]{Huberman1976striations}%
  \BibitemOpen
  \bibfield  {author} {\bibinfo {author} {\bibfnamefont {B.~A.}\ \bibnamefont
  {Huberman}},\ }\bibfield  {title} {\bibinfo {title} {Striations in chemical
  reactions},\ }\href@noop {} {\bibfield  {journal} {\bibinfo  {journal} {The
  Journal of Chemical Physics}\ }\textbf {\bibinfo {volume} {65}},\ \bibinfo
  {pages} {2013} (\bibinfo {year} {1976})}\BibitemShut {NoStop}%
\bibitem [{\citenamefont {Glotzer}\ \emph {et~al.}(1995)\citenamefont
  {Glotzer}, \citenamefont {Di~Marzio},\ and\ \citenamefont
  {Muthukumar}}]{Glotzer1995reaction}%
  \BibitemOpen
  \bibfield  {author} {\bibinfo {author} {\bibfnamefont {S.~C.}\ \bibnamefont
  {Glotzer}}, \bibinfo {author} {\bibfnamefont {E.~A.}\ \bibnamefont
  {Di~Marzio}},\ and\ \bibinfo {author} {\bibfnamefont {M.}~\bibnamefont
  {Muthukumar}},\ }\bibfield  {title} {\bibinfo {title} {Reaction-controlled
  morphology of phase-separating mixtures},\ }\href@noop {} {\bibfield
  {journal} {\bibinfo  {journal} {Physical Review Letters}\ }\textbf {\bibinfo
  {volume} {74}},\ \bibinfo {pages} {2034} (\bibinfo {year}
  {1995})}\BibitemShut {NoStop}%
\bibitem [{\citenamefont {Li}\ and\ \citenamefont {Cates}(2020)}]{Li2020noneq}%
  \BibitemOpen
  \bibfield  {author} {\bibinfo {author} {\bibfnamefont {Y.~I.}\ \bibnamefont
  {Li}}\ and\ \bibinfo {author} {\bibfnamefont {M.~E.}\ \bibnamefont {Cates}},\
  }\bibfield  {title} {\bibinfo {title} {Non-equilibrium phase separation with
  reactions: a canonical model and its behaviour},\ }\href@noop {} {\bibfield
  {journal} {\bibinfo  {journal} {Journal of Statistical Mechanics: Theory and
  Experiment}\ }\textbf {\bibinfo {volume} {2020}},\ \bibinfo {pages} {053206}
  (\bibinfo {year} {2020})}\BibitemShut {NoStop}%
\bibitem [{\citenamefont {Wurtz}\ and\ \citenamefont
  {Lee}(2018)}]{FanLee2018CRCPhase}%
  \BibitemOpen
  \bibfield  {author} {\bibinfo {author} {\bibfnamefont {J.~D.}\ \bibnamefont
  {Wurtz}}\ and\ \bibinfo {author} {\bibfnamefont {C.~F.}\ \bibnamefont
  {Lee}},\ }\bibfield  {title} {\bibinfo {title} {Chemical-reaction-controlled
  phase separated drops: Formation, size selection, and coarsening},\
  }\href@noop {} {\bibfield  {journal} {\bibinfo  {journal} {Physical Review
  Letters}\ }\textbf {\bibinfo {volume} {120}},\ \bibinfo {pages} {078102}
  (\bibinfo {year} {2018})}\BibitemShut {NoStop}%
\bibitem [{\citenamefont {Carati}\ and\ \citenamefont
  {Lefever}(1997)}]{LefeverCarati1997}%
  \BibitemOpen
  \bibfield  {author} {\bibinfo {author} {\bibfnamefont {D.}~\bibnamefont
  {Carati}}\ and\ \bibinfo {author} {\bibfnamefont {R.}~\bibnamefont
  {Lefever}},\ }\bibfield  {title} {\bibinfo {title} {Chemical freezing of
  phase separation in immiscible binary mixtures},\ }\href@noop {} {\bibfield
  {journal} {\bibinfo  {journal} {Physical Review E}\ }\textbf {\bibinfo
  {volume} {56}},\ \bibinfo {pages} {3127} (\bibinfo {year}
  {1997})}\BibitemShut {NoStop}%
\bibitem [{\citenamefont {Avanzini}\ \emph {et~al.}(2021)\citenamefont
  {Avanzini}, \citenamefont {Penocchio}, \citenamefont {Falasco},\ and\
  \citenamefont {Esposito}}]{Esposito2021NonidealCRN}%
  \BibitemOpen
  \bibfield  {author} {\bibinfo {author} {\bibfnamefont {F.}~\bibnamefont
  {Avanzini}}, \bibinfo {author} {\bibfnamefont {E.}~\bibnamefont {Penocchio}},
  \bibinfo {author} {\bibfnamefont {G.}~\bibnamefont {Falasco}},\ and\ \bibinfo
  {author} {\bibfnamefont {M.}~\bibnamefont {Esposito}},\ }\bibfield  {title}
  {\bibinfo {title} {Nonequilibrium thermodynamics of non-ideal chemical
  reaction networks},\ }\href@noop {} {\bibfield  {journal} {\bibinfo
  {journal} {The Journal of Chemical Physics}\ }\textbf {\bibinfo {volume}
  {154}},\ \bibinfo {pages} {094114} (\bibinfo {year} {2021})}\BibitemShut
  {NoStop}%
\bibitem [{\citenamefont {Bauermann}\ \emph {et~al.}(2021)\citenamefont
  {Bauermann}, \citenamefont {Laha}, \citenamefont {McCall}, \citenamefont
  {J{\"u}licher},\ and\ \citenamefont {Weber}}]{bauermann2021chemical}%
  \BibitemOpen
  \bibfield  {author} {\bibinfo {author} {\bibfnamefont {J.}~\bibnamefont
  {Bauermann}}, \bibinfo {author} {\bibfnamefont {S.}~\bibnamefont {Laha}},
  \bibinfo {author} {\bibfnamefont {P.~M.}\ \bibnamefont {McCall}}, \bibinfo
  {author} {\bibfnamefont {F.}~\bibnamefont {J{\"u}licher}},\ and\ \bibinfo
  {author} {\bibfnamefont {C.~A.}\ \bibnamefont {Weber}},\ }\bibfield  {title}
  {\bibinfo {title} {Chemical kinetics and mass action in coexisting phases},\
  }\href@noop {} {\bibfield  {journal} {\bibinfo  {journal} {arXiv preprint
  arXiv:2112.07576}\ } (\bibinfo {year} {2021})}\BibitemShut {NoStop}%
\bibitem [{\citenamefont {Bazant}(2013)}]{Bazant2013}%
  \BibitemOpen
  \bibfield  {author} {\bibinfo {author} {\bibfnamefont {M.~Z.}\ \bibnamefont
  {Bazant}},\ }\bibfield  {title} {\bibinfo {title} {Theory of chemical
  kinetics and charge transfer based on nonequilibrium thermodynamics},\
  }\href@noop {} {\bibfield  {journal} {\bibinfo  {journal} {Accounts of
  Chemical Research}\ }\textbf {\bibinfo {volume} {46}},\ \bibinfo {pages}
  {1144} (\bibinfo {year} {2013})}\BibitemShut {NoStop}%
\bibitem [{\citenamefont {Kirschbaum}\ and\ \citenamefont
  {Zwicker}(2021)}]{Zwicker2021controlling}%
  \BibitemOpen
  \bibfield  {author} {\bibinfo {author} {\bibfnamefont {J.}~\bibnamefont
  {Kirschbaum}}\ and\ \bibinfo {author} {\bibfnamefont {D.}~\bibnamefont
  {Zwicker}},\ }\bibfield  {title} {\bibinfo {title} {Controlling biomolecular
  condensates via chemical reactions},\ }\href@noop {} {\bibfield  {journal}
  {\bibinfo  {journal} {Journal of The Royal Society Interface}\ }\textbf
  {\bibinfo {volume} {18}},\ \bibinfo {pages} {20210255} (\bibinfo {year}
  {2021})}\BibitemShut {NoStop}%
\bibitem [{\citenamefont {Gillespie}(1992)}]{Gillespie1992rigurous}%
  \BibitemOpen
  \bibfield  {author} {\bibinfo {author} {\bibfnamefont {D.~T.}\ \bibnamefont
  {Gillespie}},\ }\bibfield  {title} {\bibinfo {title} {A rigorous derivation
  of the chemical master equation},\ }\href@noop {} {\bibfield  {journal}
  {\bibinfo  {journal} {Physica A: Statistical Mechanics and its Applications}\
  }\textbf {\bibinfo {volume} {188}},\ \bibinfo {pages} {404} (\bibinfo {year}
  {1992})}\BibitemShut {NoStop}%
\bibitem [{\citenamefont {Schnoerr}\ \emph {et~al.}(2017)\citenamefont
  {Schnoerr}, \citenamefont {Sanguinetti},\ and\ \citenamefont
  {Grima}}]{Grima2017Approximation}%
  \BibitemOpen
  \bibfield  {author} {\bibinfo {author} {\bibfnamefont {D.}~\bibnamefont
  {Schnoerr}}, \bibinfo {author} {\bibfnamefont {G.}~\bibnamefont
  {Sanguinetti}},\ and\ \bibinfo {author} {\bibfnamefont {R.}~\bibnamefont
  {Grima}},\ }\bibfield  {title} {\bibinfo {title} {Approximation and inference
  methods for stochastic biochemical kinetics—a tutorial review},\
  }\href@noop {} {\bibfield  {journal} {\bibinfo  {journal} {Journal of Physics
  A: Mathematical and Theoretical}\ }\textbf {\bibinfo {volume} {50}},\
  \bibinfo {pages} {093001} (\bibinfo {year} {2017})}\BibitemShut {NoStop}%
\bibitem [{\citenamefont {Horn}\ and\ \citenamefont
  {Jackson}(1972)}]{Horn1972}%
  \BibitemOpen
  \bibfield  {author} {\bibinfo {author} {\bibfnamefont {F.}~\bibnamefont
  {Horn}}\ and\ \bibinfo {author} {\bibfnamefont {R.}~\bibnamefont {Jackson}},\
  }\bibfield  {title} {\bibinfo {title} {General mass action kinetics},\
  }\href@noop {} {\bibfield  {journal} {\bibinfo  {journal} {Archive for
  rational mechanics and analysis}\ }\textbf {\bibinfo {volume} {47}},\
  \bibinfo {pages} {81} (\bibinfo {year} {1972})}\BibitemShut {NoStop}%
\bibitem [{\citenamefont {Feinberg}(1972)}]{feinberg1972complex}%
  \BibitemOpen
  \bibfield  {author} {\bibinfo {author} {\bibfnamefont {M.}~\bibnamefont
  {Feinberg}},\ }\bibfield  {title} {\bibinfo {title} {Complex balancing in
  general kinetic systems},\ }\href@noop {} {\bibfield  {journal} {\bibinfo
  {journal} {Archive for rational mechanics and analysis}\ }\textbf {\bibinfo
  {volume} {49}},\ \bibinfo {pages} {187} (\bibinfo {year} {1972})}\BibitemShut
  {NoStop}%
\bibitem [{\citenamefont {Feinberg}(1995)}]{Feinberg1995}%
  \BibitemOpen
  \bibfield  {author} {\bibinfo {author} {\bibfnamefont {M.}~\bibnamefont
  {Feinberg}},\ }\bibfield  {title} {\bibinfo {title} {The existence and
  uniqueness of steady states for a class of chemical reaction networks},\
  }\href@noop {} {\bibfield  {journal} {\bibinfo  {journal} {Archive for
  Rational Mechanics and Analysis}\ }\textbf {\bibinfo {volume} {132}},\
  \bibinfo {pages} {311} (\bibinfo {year} {1995})}\BibitemShut {NoStop}%
\bibitem [{\citenamefont {Anderson}\ \emph {et~al.}(2010)\citenamefont
  {Anderson}, \citenamefont {Craciun},\ and\ \citenamefont
  {Kurtz}}]{Anderson2010ProductFormCRN}%
  \BibitemOpen
  \bibfield  {author} {\bibinfo {author} {\bibfnamefont {D.~F.}\ \bibnamefont
  {Anderson}}, \bibinfo {author} {\bibfnamefont {G.}~\bibnamefont {Craciun}},\
  and\ \bibinfo {author} {\bibfnamefont {T.~G.}\ \bibnamefont {Kurtz}},\
  }\bibfield  {title} {\bibinfo {title} {Product-form stationary distributions
  for deficiency zero chemical reaction networks},\ }\href@noop {} {\bibfield
  {journal} {\bibinfo  {journal} {Bulletin of Mathematical Biology}\ }\textbf
  {\bibinfo {volume} {72}},\ \bibinfo {pages} {1947} (\bibinfo {year}
  {2010})}\BibitemShut {NoStop}%
\bibitem [{\citenamefont {Polettini}\ \emph {et~al.}(2015)\citenamefont
  {Polettini}, \citenamefont {Wachtel},\ and\ \citenamefont
  {Esposito}}]{Polettini2015Deficiency}%
  \BibitemOpen
  \bibfield  {author} {\bibinfo {author} {\bibfnamefont {M.}~\bibnamefont
  {Polettini}}, \bibinfo {author} {\bibfnamefont {A.}~\bibnamefont {Wachtel}},\
  and\ \bibinfo {author} {\bibfnamefont {M.}~\bibnamefont {Esposito}},\
  }\bibfield  {title} {\bibinfo {title} {Dissipation in noisy chemical
  networks: The role of deficiency},\ }\href@noop {} {\bibfield  {journal}
  {\bibinfo  {journal} {The Journal of Chemical Physics}\ }\textbf {\bibinfo
  {volume} {143}},\ \bibinfo {pages} {184103} (\bibinfo {year}
  {2015})}\BibitemShut {NoStop}%
\bibitem [{\citenamefont {Rao}\ and\ \citenamefont
  {Esposito}(2018)}]{Rao2018Conservation}%
  \BibitemOpen
  \bibfield  {author} {\bibinfo {author} {\bibfnamefont {R.}~\bibnamefont
  {Rao}}\ and\ \bibinfo {author} {\bibfnamefont {M.}~\bibnamefont {Esposito}},\
  }\bibfield  {title} {\bibinfo {title} {Conservation laws and work fluctuation
  relations in chemical reaction networks},\ }\href@noop {} {\bibfield
  {journal} {\bibinfo  {journal} {The Journal of Chemical Physics}\ }\textbf
  {\bibinfo {volume} {149}},\ \bibinfo {pages} {245101} (\bibinfo {year}
  {2018})}\BibitemShut {NoStop}%
\bibitem [{\citenamefont {H\"anggi}\ \emph {et~al.}(1990)\citenamefont
  {H\"anggi}, \citenamefont {Talkner},\ and\ \citenamefont
  {Borkovec}}]{HanggiReactionRate}%
  \BibitemOpen
  \bibfield  {author} {\bibinfo {author} {\bibfnamefont {P.}~\bibnamefont
  {H\"anggi}}, \bibinfo {author} {\bibfnamefont {P.}~\bibnamefont {Talkner}},\
  and\ \bibinfo {author} {\bibfnamefont {M.}~\bibnamefont {Borkovec}},\
  }\bibfield  {title} {\bibinfo {title} {Reaction-rate theory: fifty years
  after {K}ramers},\ }\href@noop {} {\bibfield  {journal} {\bibinfo  {journal}
  {Reviews of Modern Physics}\ }\textbf {\bibinfo {volume} {62}},\ \bibinfo
  {pages} {251} (\bibinfo {year} {1990})}\BibitemShut {NoStop}%
\bibitem [{\citenamefont {Kondepundi}\ and\ \citenamefont
  {Prigogine}(2014)}]{prigogineThermodynamics}%
  \BibitemOpen
  \bibfield  {author} {\bibinfo {author} {\bibfnamefont {D.}~\bibnamefont
  {Kondepundi}}\ and\ \bibinfo {author} {\bibfnamefont {I.}~\bibnamefont
  {Prigogine}},\ }\href@noop {} {\emph {\bibinfo {title} {Modern
  Thermodynamics}}}\ (\bibinfo  {publisher} {John Wiley \& Sons Ltd},\ \bibinfo
  {address} {New York},\ \bibinfo {year} {2014})\BibitemShut {NoStop}%
\bibitem [{\citenamefont {J{\"u}licher}\ \emph {et~al.}(1997)\citenamefont
  {J{\"u}licher}, \citenamefont {Ajdari},\ and\ \citenamefont
  {Prost}}]{julicher1997motors}%
  \BibitemOpen
  \bibfield  {author} {\bibinfo {author} {\bibfnamefont {F.}~\bibnamefont
  {J{\"u}licher}}, \bibinfo {author} {\bibfnamefont {A.}~\bibnamefont
  {Ajdari}},\ and\ \bibinfo {author} {\bibfnamefont {J.}~\bibnamefont
  {Prost}},\ }\bibfield  {title} {\bibinfo {title} {Modeling molecular
  motors},\ }\href@noop {} {\bibfield  {journal} {\bibinfo  {journal} {Reviews
  of Modern Physics}\ }\textbf {\bibinfo {volume} {69}},\ \bibinfo {pages}
  {1269} (\bibinfo {year} {1997})}\BibitemShut {NoStop}%
\bibitem [{\citenamefont {Schnakenberg}(1976)}]{Schnakenberg1976Network}%
  \BibitemOpen
  \bibfield  {author} {\bibinfo {author} {\bibfnamefont {J.}~\bibnamefont
  {Schnakenberg}},\ }\bibfield  {title} {\bibinfo {title} {Network theory of
  microscopic and macroscopic behavior of master equation systems},\
  }\href@noop {} {\bibfield  {journal} {\bibinfo  {journal} {Reviews of Modern
  Physics}\ }\textbf {\bibinfo {volume} {48}},\ \bibinfo {pages} {571}
  (\bibinfo {year} {1976})}\BibitemShut {NoStop}%
\bibitem [{\citenamefont {Anderson}\ and\ \citenamefont
  {Kurtz}(2015)}]{anderson2015stochastic}%
  \BibitemOpen
  \bibfield  {author} {\bibinfo {author} {\bibfnamefont {D.~F.}\ \bibnamefont
  {Anderson}}\ and\ \bibinfo {author} {\bibfnamefont {T.~G.}\ \bibnamefont
  {Kurtz}},\ }\href@noop {} {\emph {\bibinfo {title} {Stochastic analysis of
  biochemical systems}}},\ Vol.\ \bibinfo {volume} {674}\ (\bibinfo
  {publisher} {Springer},\ \bibinfo {address} {Zurich},\ \bibinfo {year}
  {2015})\BibitemShut {NoStop}%
\bibitem [{\citenamefont {Rao}\ and\ \citenamefont
  {Esposito}(2016)}]{Esposito2016noneqCRN}%
  \BibitemOpen
  \bibfield  {author} {\bibinfo {author} {\bibfnamefont {R.}~\bibnamefont
  {Rao}}\ and\ \bibinfo {author} {\bibfnamefont {M.}~\bibnamefont {Esposito}},\
  }\bibfield  {title} {\bibinfo {title} {Nonequilibrium thermodynamics of
  chemical reaction networks: Wisdom from stochastic thermodynamics},\
  }\href@noop {} {\bibfield  {journal} {\bibinfo  {journal} {Physical Review
  X}\ }\textbf {\bibinfo {volume} {6}},\ \bibinfo {pages} {041064} (\bibinfo
  {year} {2016})}\BibitemShut {NoStop}%
\bibitem [{\citenamefont {Gillespie}(1977)}]{Gillespie1977algorithm}%
  \BibitemOpen
  \bibfield  {author} {\bibinfo {author} {\bibfnamefont {D.~T.}\ \bibnamefont
  {Gillespie}},\ }\bibfield  {title} {\bibinfo {title} {Exact stochastic
  simulation of coupled chemical reactions},\ }\href@noop {} {\bibfield
  {journal} {\bibinfo  {journal} {The Journal of Physical Chemistry}\ }\textbf
  {\bibinfo {volume} {81}},\ \bibinfo {pages} {2340} (\bibinfo {year}
  {1977})}\BibitemShut {NoStop}%
\bibitem [{\citenamefont {Gang}(1986)}]{gang1986lyapounov}%
  \BibitemOpen
  \bibfield  {author} {\bibinfo {author} {\bibfnamefont {H.}~\bibnamefont
  {Gang}},\ }\bibfield  {title} {\bibinfo {title} {Lyapounov function and
  stationary probability distributions},\ }\href@noop {} {\bibfield  {journal}
  {\bibinfo  {journal} {Zeitschrift f{\"u}r Physik B Condensed Matter}\
  }\textbf {\bibinfo {volume} {65}},\ \bibinfo {pages} {103} (\bibinfo {year}
  {1986})}\BibitemShut {NoStop}%
\bibitem [{\citenamefont {Ge}\ and\ \citenamefont
  {Qian}(2017)}]{ge2017mathematical}%
  \BibitemOpen
  \bibfield  {author} {\bibinfo {author} {\bibfnamefont {H.}~\bibnamefont
  {Ge}}\ and\ \bibinfo {author} {\bibfnamefont {H.}~\bibnamefont {Qian}},\
  }\bibfield  {title} {\bibinfo {title} {Mathematical formalism of
  nonequilibrium thermodynamics for nonlinear chemical reaction systems with
  general rate law},\ }\href@noop {} {\bibfield  {journal} {\bibinfo  {journal}
  {Journal of Statistical Physics}\ }\textbf {\bibinfo {volume} {166}},\
  \bibinfo {pages} {190} (\bibinfo {year} {2017})}\BibitemShut {NoStop}%
\bibitem [{\citenamefont {Anderson}\ \emph {et~al.}(2015)\citenamefont
  {Anderson}, \citenamefont {Craciun}, \citenamefont {Gopalkrishnan},\ and\
  \citenamefont {Wiuf}}]{Anderson2015Lyapunov}%
  \BibitemOpen
  \bibfield  {author} {\bibinfo {author} {\bibfnamefont {D.~F.}\ \bibnamefont
  {Anderson}}, \bibinfo {author} {\bibfnamefont {G.}~\bibnamefont {Craciun}},
  \bibinfo {author} {\bibfnamefont {M.}~\bibnamefont {Gopalkrishnan}},\ and\
  \bibinfo {author} {\bibfnamefont {C.}~\bibnamefont {Wiuf}},\ }\bibfield
  {title} {\bibinfo {title} {Lyapunov functions, stationary distributions, and
  non-equilibrium potential for reaction networks},\ }\href@noop {} {\bibfield
  {journal} {\bibinfo  {journal} {Bulletin of Mathematical Biology}\ }\textbf
  {\bibinfo {volume} {77}},\ \bibinfo {pages} {1744} (\bibinfo {year}
  {2015})}\BibitemShut {NoStop}%
\bibitem [{\citenamefont {Anderson}\ and\ \citenamefont
  {Nguyen}(2019)}]{Anderson2019Results}%
  \BibitemOpen
  \bibfield  {author} {\bibinfo {author} {\bibfnamefont {D.~F.}\ \bibnamefont
  {Anderson}}\ and\ \bibinfo {author} {\bibfnamefont {T.~D.}\ \bibnamefont
  {Nguyen}},\ }\bibfield  {title} {\bibinfo {title} {Results on stochastic
  reaction networks with non-mass action kinetics},\ }\href@noop {} {\bibfield
  {journal} {\bibinfo  {journal} {Mathematical Biosciences and Engineering}\
  }\textbf {\bibinfo {volume} {16}},\ \bibinfo {pages} {2118} (\bibinfo {year}
  {2019})}\BibitemShut {NoStop}%
\bibitem [{\citenamefont {De~Groot}\ and\ \citenamefont
  {Mazur}(2013)}]{groot&mazur}%
  \BibitemOpen
  \bibfield  {author} {\bibinfo {author} {\bibfnamefont {S.}~\bibnamefont
  {De~Groot}}\ and\ \bibinfo {author} {\bibfnamefont {P.}~\bibnamefont
  {Mazur}},\ }\href@noop {} {\emph {\bibinfo {title} {Non-Equilibrium
  Thermodynamics}}},\ Dover Books on Physics\ (\bibinfo  {publisher} {Dover
  Publications},\ \bibinfo {address} {Amsterdam},\ \bibinfo {year}
  {2013})\BibitemShut {NoStop}%
\bibitem [{\citenamefont {Onsager}(1931)}]{onsager1931reciprocal}%
  \BibitemOpen
  \bibfield  {author} {\bibinfo {author} {\bibfnamefont {L.}~\bibnamefont
  {Onsager}},\ }\bibfield  {title} {\bibinfo {title} {Reciprocal relations in
  irreversible processes. {I}.},\ }\href@noop {} {\bibfield  {journal}
  {\bibinfo  {journal} {Physical Review}\ }\textbf {\bibinfo {volume} {37}},\
  \bibinfo {pages} {405} (\bibinfo {year} {1931})}\BibitemShut {NoStop}%
\bibitem [{\citenamefont {Cahn}\ and\ \citenamefont
  {Hilliard}(1958)}]{cahn1958}%
  \BibitemOpen
  \bibfield  {author} {\bibinfo {author} {\bibfnamefont {J.~W.}\ \bibnamefont
  {Cahn}}\ and\ \bibinfo {author} {\bibfnamefont {J.~E.}\ \bibnamefont
  {Hilliard}},\ }\bibfield  {title} {\bibinfo {title} {Free energy of a
  nonuniform system. {I}. {I}nterfacial free energy},\ }\href@noop {}
  {\bibfield  {journal} {\bibinfo  {journal} {Journal of Chemical Physics}\
  }\textbf {\bibinfo {volume} {28}},\ \bibinfo {pages} {258} (\bibinfo {year}
  {1958})}\BibitemShut {NoStop}%
\bibitem [{\citenamefont {Hondele}\ \emph {et~al.}(2020)\citenamefont
  {Hondele}, \citenamefont {Heinrich}, \citenamefont {De~Los~Rios},\ and\
  \citenamefont {Weis}}]{Weis2020Phasing}%
  \BibitemOpen
  \bibfield  {author} {\bibinfo {author} {\bibfnamefont {M.}~\bibnamefont
  {Hondele}}, \bibinfo {author} {\bibfnamefont {S.}~\bibnamefont {Heinrich}},
  \bibinfo {author} {\bibfnamefont {P.}~\bibnamefont {De~Los~Rios}},\ and\
  \bibinfo {author} {\bibfnamefont {K.}~\bibnamefont {Weis}},\ }\bibfield
  {title} {\bibinfo {title} {{Membraneless organelles: phasing out of
  equilibrium}},\ }\href@noop {} {\bibfield  {journal} {\bibinfo  {journal}
  {Emerging Topics in Life Sciences}\ }\textbf {\bibinfo {volume} {4}},\
  \bibinfo {pages} {343} (\bibinfo {year} {2020})}\BibitemShut {NoStop}%
\bibitem [{\citenamefont {Mehta}\ \emph {et~al.}(2016)\citenamefont {Mehta},
  \citenamefont {Lang},\ and\ \citenamefont {Schwab}}]{Mehta2016}%
  \BibitemOpen
  \bibfield  {author} {\bibinfo {author} {\bibfnamefont {P.}~\bibnamefont
  {Mehta}}, \bibinfo {author} {\bibfnamefont {A.~H.}\ \bibnamefont {Lang}},\
  and\ \bibinfo {author} {\bibfnamefont {D.~J.}\ \bibnamefont {Schwab}},\
  }\bibfield  {title} {\bibinfo {title} {Landauer in the age of synthetic
  biology: Energy consumption and information processing in biochemical
  networks},\ }\href@noop {} {\bibfield  {journal} {\bibinfo  {journal}
  {Journal of Statistical Physics}\ }\textbf {\bibinfo {volume} {162}},\
  \bibinfo {pages} {1153} (\bibinfo {year} {2016})}\BibitemShut {NoStop}%
\end{thebibliography}%

\end{document}